\documentclass[12pt]{cernart}
\usepackage{graphicx}
\usepackage{here}
\begin{document}
\begin{titlepage}
\docnum{CERN--EP/2002--038}
\date{22 February 2002}

\vglue.5cm

\title{OBSERVATIONAL EVIDENCE FOR GRAVITATIONALLY \\
 TRAPPED MASSIVE AXION(-LIKE) PARTICLES}
   
\begin{Authlist}
L. DiLella\IAref{a1}{*} and K.~Zioutas\IIref{a1}{a2}
\end{Authlist}

\vglue1cm

\begin{abstract}
Several unexpected astrophysical observations can be explained by 
gravitationally captured massive axions or axion-like particles, which are
produced  inside the Sun  or other stars and are accumulated over cosmic times. 
Their radiative decay  in  solar outer space 
would give rise to a  `self-irradiation' of the  whole star, 
providing the time-independent component of the corona heating source
(we do not address here the flaring Sun).
In analogy with the Sun-irradiated Earth atmosphere, the temperature and density 
gradient in the corona$-$chromosphere transition region is suggestive for 
an omnipresent irradiation of the Sun, which is the strongest evidence
for the generic axion-like scenario. The same mechanism is compatible with 
phenomena like the solar wind, the X-rays from the dark-side of the Moon, 
the  X-Ray  Background Radiation,  the diffuse X-ray excesses (below $\sim 1$ 
keV),  the non-cooling of oldest Stars, etc. 
A temperature of $\sim 10^6$ K  is observed in various places,  
while the radiative decay of  a population of such elusive 
particles mimics a hot gas, which fits unexpected astrophysical X-ray
observations.
Furthermore, the recently reconstructed quiet solar X-ray spectrum  during 
solar minimum supports this work, since it covers the expected energy range, 
and  it is consistent with the result of a simulation based on  Kaluza-Klein 
axions above $\sim$ 1 keV. 
The derived axion luminosity ($L_a\approx 0.16L_{\odot}$) fits the cosmic
energy density spectrum and is compatible within 2$\sigma $ with the recent 
SNO result, showing the important interplay between any exotic energy loss 
mechanism and neutrino production.
At lower energies, using also a ROSAT observation,
only $\sim 3\%$ of the X-ray intensity is explained.
Data from orbiting X-ray Telescopes provide upper limits for 
particle decay rates 1 AU from the Sun, and suggest  new types of 
searches on Earth or in space. In particular,
X-ray observatories, with an unrivalled equivalent fiducial volume of 
$\sim 10^3~m^3$ for the 0.1  - 10 keV range, can search  for the radiative 
decay of new particles even from existing data.
This work introduces the elongation angle of the X-ray Telescope relative to the
Sun as a relevant new parameter.

\end{abstract}

\vglue1cm
\submitted{(To appear in  Astroparticle  Physics) }
  
\Instfoot{a1}{CERN, CH-1211 Geneva 23, Switzerland.}
\Instfoot{a2}{Physics Department, University of
Thessaloniki, 
  GR 541 24  Thessaloniki, Greece.}
\Anotfoot{*}{E-mails: Luigi.Di.Lella@cern.ch,~Konstantin.Zioutas@cern.ch} 

\end{titlepage}

\section{INTRODUCTION}

The direct detection of dark-matter (DM) particles has proved elusive since
the first gravitational observation of  non-luminous matter in
the Universe. 
So far, the outcome of the intense experimental and theoretical  work 
in the field of dark matter during the last $\sim 20$ years was the birth 
of the new discipline of astroparticle physics.

In this work, a rather old question is reconsidered as to whether a large number of
as yet unexplained astrophysical phenomena occur because of the involvement of 
novel very weakly interacting particles or additional as yet unknown properties
of existing particles.
For example, in order to explain the ionization  state of the intergalactic
medium, and the anomalous ionization of the interstellar medium
\cite{bowyerxxx},
speculations included a widespread source of ionizing UV-photons from 
the electromagnetic decay  of real or hypothetical exotic particles (clustered
in haloes)
\cite{bowyer},
e.g., massive dark-matter neutrinos
\cite{sciama,abazajian};
however, observations appear to have ruled out this model
\cite{bowyerxxx,bowyer1xxx}.
Considerations based on axions apply as well,  invoking either its $2\gamma $ 
decay mode or its coherent conversion to a photon inside
astrophysical electric/magnetic fields via the Primakoff effect
\footnote{
a) The absence of
a monochromatic axion line from the night Sky expected from the 
$a\rightarrow \gamma \gamma$ decay  of relic axions in the visible
\cite{turner}
almost excluded an axion rest mass in the $\sim  1$--10~eV range. 
b) Until recently, a conventionally expected
thermal X-ray spectrum from solar axions converted inside the solar magnetic 
fields with mean energy of $\sim 4.4$~keV could not have  been disentangled 
from the derived solar X-ray spectrum (e.g. Ref.~\cite{gudel3xxx}).
In any case, such a mechanism could not explain the  main observations addressed
in this work, because such converted axions give rise to photons always emitted 
away from the Sun. Moreover, even in the scenario of massive axions from the
Sun, the possibility to detect photons emitted in flight through axion decays
was considered to be "presumably indistinguishable from the general background
radiation"
\cite{dienes}.
}.

Recently, in order to explain the as yet unknown underlying mechanism(s) of 
the Gamma Ray Bursts, massive axions with properties far beyond the widely 
accepted theoretical axion concepts have been considered, providing a built-in
dissipationless energy transfer mechanism from the hypothetical energy 
generating core to the outside layers  some 100$-$1000 km away
\cite{grb}.
Even though none of such intriguing ideas has been established so far
\cite{henry1},
an additional electromagnetic energy source in the cosmos seems to be
necessary.
Structure formation in warm dark matter  (WDM)
cosmological models
\cite{narayanan}
provides a lower limit to the mass of the WDM particle candidate of 
$\sim 0.75$ keV, but with a radiative decay lifetime as large as 
$\sim 10^{16}\times $Hubble time;
however, halos in Galaxies and clusters of Galaxies can be an enormous 
"fiducial volume" of DM particles.

In this work, we argue that the photon emission of some hypothetical 
particles, referred to more generally  as `axion-like', could be 
involved  in different unexplained astrophysical observations. 
An extensive search of astrophysical literature has been undertaken, 
which includes also some 10--50 years old observations. In this work we 
focus mainly  on :
\begin{itemize}
\item[a)] the solar corona problem and related observations;
\item[b)] the  observed X-rays from the direction of the dark side 
of the Moon;
\item[c)] the soft-X-ray background radiation;
\item[d)] the (diffuse) soft X-ray excess phenomenon;
\item[e)] first simulation results  in the frame of an axion scenario.
\end{itemize}
Following the reasoning of this work, we also suggest performing a 
{\it specific} axion search in a new type of experiment, either on the surface 
(and in space) or underground, aiming  directly at detecting the $2\gamma $ 
decay / interaction mode. The alternative case of a single photon emission 
seems to be beyond the present sensitivity of an Earth-bound detector and only 
orbiting X-ray telescopes could be considered.
In addition, some astrophysical measurements could be reconsidered or 
re-analysed.  X-ray space detectors could operate also as sensitive orbiting 
axion telescopes.

\section{OBSERVATIONAL EVIDENCE}

In this section we present some not yet understood astrophysical phenomena
or experimental results, which can be  explained  within
the frame of  the same axion-like scenario.

\subsection{Solar corona}
The existence of the solar corona has been known for more than 100 years.
However,  solar X-rays have been measured only over the last 50 years, 
providing an unexpected and anomalously high temperature
\cite{schmitt,xrays,podlaxxx}.
The corona is the only atmospheric layer of the Sun that emits almost thermally
in X-rays
\cite{parnell}.
The average temperature of the quiet-Sun  corona is
$\sim 2\cdot 10^6$~K.
Although its phenomenology is well understood,
the coronal heating remains one of the most puzzling problems in solar physics
\cite{schmitt,podlaxxx,orlando,pevtsovyyy,nandraxxx,
dmitruk7xxx,benz9xxx,fluda,benz,feldmanX,gperes9zzz}.
Recently, reconstructed X-ray energy spectra have been published, 
providing additional valuable information about our Sun, such as temperature, 
solar cycle dynamics, etc.
\cite{orlando}.
The quiet-Sun X-ray luminosity  
represents only a fraction of $\leq 10^{-7}$ of the total solar luminosity
($L_{\odot}^x \approx 2\cdot 10^{27}$ erg/s at solar maximum
\cite{schmitt5xxx,haisch,birker},
which never drops to zero). 
Therefore, energy balance problems are irrelevant
\footnote{
In this work we refer to the quiet Sun, where also nano-flares  occur 
\cite{apr333x}
and they could account for a significant fraction of the coronal heating
\cite{apr}. 
The X-ray power of the 
active Sun is higher by a factor of  $\sim 20$, while
its temperature of $\approx  8$--20~MK \cite{orlando,birker} 
is equal to that of the Sun core.
}.
Thus, the main puzzles with the solar corona {\sf and} the very thin
\cite{doschek9zzz} 
transition  region (TR) between the corona and the underlying 
chromosphere (the least understood region of the solar atmosphere
\cite{judgeyyy})
are the following :
\begin{itemize}
\item[a)] 
The hot corona cannot be in equilibrium with the 
$\sim 300$ times cooler solar surface  underneath (violating thus 
the second law of thermodynamics
\cite{wolfsonyyy}), 
which emits an almost perfect blackbody radiation in the visible
\cite{lean}
(Fig.~1b).
\item[b)] In order to maintain the quiet Sun high temperature corona, some
{\it nonthermally} supplied energy must be dissipated in the upper atmosphere
\cite{alexander},
which is lost as a large inward heat flux into the transition region, solar
wind energy and transition region pressure
\cite{wolfson5xxx,evje7xxx}; 
these processes are  thermodynamically allowed, because of the
purported coronal temperature increase with height.
\item[c)] One must explain the abrupt (in some models 
\cite{warren9zzz}
even within $\sim 100$ km) temperature increase 
(from $\sim 8\cdot 10^3$ K to $\sim 5\cdot 10^5$ K)
in the chromosphere/corona transition region (Fig.~1a), against physical 
expectation
\cite{maran,phillips,ayresZZZ,morphology}
\footnote{
Note that most of the solar light comes from the
$\sim 100$ km visible  photosphere ($\sim 5800$ K). Above that, there is the 
chromosphere which is  astonishingly hot (up to 25\,000~K), and above that, 
the corona (up to a few MK and locally much more
\cite{phillips}).
Corona X-rays have been observed out to $\sim 1$ solar radius. The outer
corona expands  into the interplanetary space and slowly cools off. At $\sim
1$--100~AU the temperature is still $\sim 10^5$~K 
\cite{keller}
(see  also R.F. Stein in Ref.~\cite{maran}). 
If the corona is heated by thermal processes,
it could not be  hotter than the photosphere. The second law of thermodynamics
precludes thermal transfer from the relatively cool photosphere to the much
hotter corona
\cite{wolfsonyyy}.
Therefore, it must be heated by some non-thermal process  
\cite{maran,keller},
e.g., by electromagnetic energy input
\cite{wolfsonyyy}.
Obviously,  only below a certain density  can the energy input 
(whatever the required nonthermal energy transfer mechanism is in reality)
be sufficient to increase the temperature or trigger an unexpected 
process.
}. 

\end{itemize}

\noindent
This is `the solar corona problem', which is well stated in a recent review
article 
\cite{apr} :

\noindent
{\it \sf "everything above the photosphere ..... would not be there 
at all"}.

The step-like change of the corona temperature
coincides in space with a similar (opposite) density gradient (Fig.~1a),
thus suggesting
a common origin. Qualitatively, this peculiar behaviour of the Sun atmosphere
is suggestive for some external irradiation (pressure) acting on the whole Sun,
and only such a configuration can cause  the  `compression' {\sf and}  the 
heating  of the intervening solar atmosphere (= corona region).
Depending on the energy,  these photons  are absorbed mainly at a certain depth 
(as seen from outside the Sun) due to the exponential increase of the density 
with decreasing height of the solar atmosphere (see Fig.~1a and 
Ref.~\cite{maran}).
One should keep in mind that the density at the place where both steps occur is 
$\sim 10^{-(12\pm 1)}$~g/cm$^3$, i.e. an excellent  vacuum, which actually 
does not facilitate a conventional explanation of this  observation.

The column density   in the solar atmosphere at an 
altitude of $\sim 2000$ km  down to  $\sim 1000$ km  is $\sim 10~\mu$g/cm$^2$
to $\sim 1~$mg/cm$^2$, respectively
\cite{cox7xxx}
\footnote{
Observations suggest that there is considerable roughness and variability of the
transition region 
\cite{apr,fontenla7xxx}, 
which might consist of many sharp temperature jumps in the different places,
but not always  at the same height
\cite{marsch7xxx} 
(see also ref. \cite{warren9zzz}).
Therefore, we use first these two penetration depths in our order of magnitude 
numerical calculation of the photoelectric absorption of the soft X-rays, 
in order to derive approximately the X-ray energy from the assumed 
axion decays around the Sun.
}.
We take the most abundant 11 solar elements in neutral form 
(H, He, C, N, O, Ne, Mg, Si, Ar, Ca, and Fe), in order to make a rough 
estimate of the energy of incident photons which can be absorbed at such 
depths of the solar atmosphere.
Photons with  an energy below  $\sim$ 50 to 350 eV, 
hitting the solar surface perpendicularly (i.e. at the minimum absorption 
thickness) can be absorbed at the considered  depths 
\cite{pdg}. 
The higher energetic X-rays penetrate deeper into the underlying chromosphere, 
where the onset of the anomalous  temperature rise above the photosphere 
occurs. Remarkably, the observationally reconstructed solar X-rays 
\cite{orlando}
above $\sim 1$ keV have a mean free path into the Sun atmosphere of 
$\sim 10$ mg/cm$^2$ ($E_{\gamma}\approx $1 keV) and $\sim 1$ g/cm$^2$
($E_{\gamma}\approx$ 6 to 8 keV), corresponding to heights of $\sim$ 700 km and 
$\sim$ 200 km, respectively. Interestingly, at a height of $\sim 500$
km there is a temperature minimum of $\sim 4000^o$K
\cite{judgeyyy},
while X-rays  below a few keV barely reach this place; 
a wide energy distribution  of incident X-rays  above 1 - 2  keV
\cite{orlando}
could be related to the  relatively slow increase of the chromosphere's 
temperature with height.  
Note that {\sf the coronal heating problem  cannot be solved without
including energization processes in the chromosphere and the transition region}
\cite{asch9zzz}.

All these findings and problems associated with the solar atmosphere
can be reconciled by assuming axion(-like) particles which  stream out of the 
interior of the Sun and undergo photon decay, giving rise to the occurence of  
otherwise unexpected nonthermal phenomena.
The photons from the $a\rightarrow \gamma \gamma $ decay are 
emitted isotropically, i.e. also towards the Sun, resulting to an external  
illumination of the  whole solar  atmosphere\footnote{
The bombardement of the Sun by cosmic material has been proposed
some 50 years ago, in order to explain the Sun's  dynamical behaviour, but 
this scenario was later ruled out
\cite{unsold7xxx}.
}.
In the dense  interior of the Sun the axions can only have a negligible impact 
due  to energy considerations, thus avoiding any conflict with the generally 
accepted solar model(s).
However, above some altitude, i.e. below a certain density, the thermodynamical
equilibrium starts getting disturbed because of an additional  external 
energy input, coming  ~$-$ironically$-$~  from the Sun itself.
Without taking into account this energy source, the temperature appears to 
increase to thermodynamically not allowed values across the chromosphere and in 
particular across the narrow transition region.

The striking similarity of the temperature and density dependence on the
altitude between the solar irradiation of the Earth atmosphere 
and Sun atmosphere
\cite{lean},
in particular around the transition region, 
strongly supports the  conclusion of an external illumination of the Sun 
in the UV$-$X-ray region  (compare  Fig.~1a with  Fig.~1d).
The observation made in the atmosphere of Venus 
\cite{keating} 
seems to be relevant too :  due to the solar irradiation, its nightside density 
($\rho \approx 10^{-15}$~g/cm$^3$ at a height of 170--190~km) and temperature 
increase during local day time by a factor of  10 and  30$\%$, respectively.
The photoionization rate peaks at an altitude of $\sim 140$~km
\cite{nagy}.
The planetary absorption depth of the solar radiation reflects the energy
of this radiation
\cite{lean}
(see {Fig.~1c}). Similarly, we conclude that for the case of the solar 
atmosphere the  photons emitted towards the Sun from the decay of nearby
exotic particles ($a \rightarrow \gamma \gamma$) must have an energy mainly 
around  $\sim$ 200 eV and also of  a few keV. This fits the appearance of the
abrupt TR {\sf and} the rather slowly increasing chromosphere temperature with
height
\cite{withbroexxx,haraxxx}.
Moreover, Monte Carlo simulations of X-rays irradiated stellar atmospheres
\cite{madejxxx}
and accretion disks
\cite{ballantynexxx}
predict a  similar stepwise temperature gradient as those shown in Fig.~1 for
the Earth and Sun atmospheres.

In addition, the following observation is compatible with the scenario of 
an external whole Sun X-ray irradiation. In coronal holes, where the electron
density is by a factor of 2 to 3 lower than normal quiet-Sun regions
\cite{haraxxx,li7xxx,doschek7xxx},
it was found that the thickness of the transition region  is  also larger
than in near quiet-Sun regions
\cite{wilhelm7xxx} 
(see also ref.
\cite{phillips1}
p. 154).

\vskip0.2cm
\noindent
\underline{\bf Related observations}.

\vskip0.2cm
\noindent
{\bf 1)}~~{\sf The solar wind :} 
the suggested axion scenario could also explain the apparently not so unrelated
origin of the solar wind, whose acceleration mechanism is one of the outstanding
problems in solar  physics
\cite{dwivedi5xxx}. 
The proton wind has a low velocity component ($\sim 350$ - 500 km/s) and a 
fast one (above $\sim 750$ km/s)
\cite{maran,wang7xxx,kahler8xxx}.
The high speed solar wind outflows are accelerated very close to the solar
surface
\cite{dere7xxx}.
The solar wind originates from the coronal holes, which are open magnetic field
regions  with a temperature of $\sim$ 1 MK
\cite{raju7xxx}
 (see also ref.
\cite{maran}).

We use in what follows the low X-ray energy derived above 
($E_{\gamma} \sim 200$ eV) in our attempt to reproduce main 
corona features, while the hard X-rays apply as well; the ejected 
photoelectrons basically carry away the energy of
the photon, i.e. their velocity is $v_e \leq $ 8500 km/s. 
Let the number of the photoelectrons be equal to that of the 
protons. Then, the onset of the equilibration
\cite{helfand8xxx}
between  electrons and protons results to a
proton kinetic energy of $T_p \leq 200$ eV, i.e. a proton velocity
$v_p \sim 200$ km/s.
This certainly not rigorous estimate provides a rather reasonable 
value compared to the measured
proton peak at $v_p \approx 500$ km/s
\cite{kahler8xxx}.
Note that the derived electron velocities leave  room for much higher ion
velocities, including  the high speed proton wind (up to $\sim 1200$ km/s), 
which makes only  $\sim 10^{-5}$-$10^{-6}$ of the bulk proton intensity.
At this level,  some protons might acquire more 
speed from the ambient electron gas, e.g.,  due to a  locally smaller 
proton-to-electron density ratio
\footnote{ 
Ions  merged with a  faster moving electron bunch exchange momentum-energy.
If the number of the ions is relatively small, they may reach the velocity of
the surrounding electron cloud, somehow similar to the old collective ion
accelerator concept
\cite{olson8xxx}. 
Also, the electron cooling in accelerators is based on the "heat
exchange" between the ions and the electron gas 
\cite{cern8xxx}.
},
or some other transient conditions.

\vskip0.3cm
\noindent
{\bf 2)}~~ {\sf The stellar coronae :} ~like our Sun,  many other Stars also have 
outer layers that are hotter than the underlying photosphere, giving rise to  
corona formation.
Observations suggest that the corona of the Sun and 
the  stellar coronae have common heating mechanism(s) 
\cite{schmitt,keller,gudel}.
Therefore, the solar corona problem is not just a solar problem 
\cite{schmitt,keller,gudel} 
and our conclusions based on solar observations become
rather global.
Measurements of UV and X-rays have revealed the ubiquitous presence
of solar-like transition regions and coronae
\cite{kashyap}
in outer atmospheres of completely different types of 
Stars
~$-$from the coolest M-type dwarfs (mass $\leq 0.3~M_{\odot}$)
\footnote{
Interestingly, according to ref.
\cite{haisch7xxx},
such Stars being fully convective could not sustain a dynamo which was thought
to be necessary for solar activities. And, these two experts add :
"Thumbing their nose at theory, these Stars still possess energetic coronas".}
to evolved giants$-$~  being "perhaps disappointing similar"
\cite{bowyer1xxx}.
 Thus, cool Stars without a surrounding corona either do not exist or
are very rare
\cite{schmitt5xxx}.
Therefore, the question about what heats the  coronae of the Sun or other cool
stars is one of the outstanding problems in astrophysics
\cite{kashyap,priest5xxx,schrijverxxx}.
\vskip0.3cm
\noindent
{\bf 3)}~~ {\sf The old Stars :} ~ it was  pointed out already in 1926 by 
Eddington that a Star in a steady state (= in radiative equilibrium) 
necessarily re-emits
completely the radiation falling on it
\cite{eddington7xxx}.
The incident radiation  maintains the surface layers at a temperature higher 
than they would otherwise have
\cite{milne7xxx},
and they radiate back in space the absorbed energy in addition to that which 
they would radiate in the absence of the incident radiation.
As with the suggested axion-related irradiation of the Sun, a similar 
self-irradiation of the Star by unstable and long-lived  axion-like particles 
can also take place.
Then, some related phenomena should have been observed 
 ~$-$even unintentionally$-$~  also from the  oldest astrophysical 
objects whose activity in general decreases  with age
\cite{fleming7xxx}. 
Thus, we looked for observational evidence
from the  following {\it fossil Stars} :

\noindent
{\sf a)}~ {\sf Millisecond Pulsars} with a spin-down age of $\sim 1$ to 10 Gyr;
according to the  Standard Models for  cooling, without a strong reheating 
mechanism
\cite{edelstein2xxx},
they are too old to expect any detectable thermal X-ray emission
\cite{becker2xxx}.
In one case, with the observed EUV emission, all conventional possibilities
were ruled out
\cite{edelstein2xxx}
\footnote{
Magnetic monopole  catalyzed nucleon decay was considered as
the additionally required heating source.
}.
X-ray emission of as yet unknown origin has also been detected 
from  a group of msec  Pulsars
\cite{becker3xxx}.

\noindent
{\sf b)}~ {\sf White Dwarfs }(WDs) with an age  above $\sim$ 10 Gyr, which are 
the very last evolution phase of a Star (as our $\sim 4.6$ Gyr old Sun will be 
in $\sim$ 6 Gyr).
We give here a few findings for these Earth-size objects with  mass  
M$\approx 0.6~M_{\odot}$ :

\begin{itemize}
\item[1.]
Coronal X-rays  have been detected from a WD with a temperature of 
(0.2 - 0.3)$\cdot 10^6$K
\cite{fleming1xxx}
\footnote{
If the captured population of particles around a Star is of
Kaluza-Klein axion-type
\cite{dilella5xxx}, 
then the older the object the less massive axions survive, implying thus a 
decreasing coronal temperature with time.
}.
\item[2.]
The abrupt and steep decrease of the number of WDs at very cool temperatures
(below $\sim$4000 K) cannot easily be explained as a cooling effect, i.e. WDs
have had not sufficient time to cool
\cite{kawalerxxx}.
A recent proposal 
\cite{bergeronxxx}
{\it to resolve the mystery}  shows  a lack of understanding  of the
actual physical conditions.
\item[3.]
Unexpectedly, WDs appear to be usually much younger than Pulsars in old binary  
systems, which should have the same age; 
in one case the cooling models provided 0.3 Gyr for the WD, while the spin-down 
age of the Pulsar was $\sim 7$ Gyr. 
This is suggestive for  an additional heating source, which slows down 
the cooling of the  WDs
\cite{schonberner}.
\end{itemize}

\vskip0.2cm
\noindent
Qualitatively, such diverse observations are in favour of axion-like 
particles gravitationally trapped by the Star itself.
If these particles are sufficiently long lived, a part of them  can survive 
the evolutionary  steps of the Star, giving rise to  astrophysical 
phenomena like winds, stellar coronae, X-ray/EUV emission, 
non-cooling of old Stars, etc.
The X-rays from the decay of gravitationally captured axions
can mimic thermal emission and could lead to a wrong interpretation of the black
body emission from the Star surface. For example, the recently claimed evidence
for a compact (quark) star
\cite{quarkstar} 
could be alternatively explained by the mechanism suggested in
this work.

\subsection{X-rays from the direction of the dark side of the Moon}

We {reconsider} here one lunar experiment performed by the ROSAT 
\cite{trumperXX}
orbiting X-ray telescope in 1990,
which provided the first X-ray `photo' (Fig.~2) from the sunlit side of the Moon
\cite{moon1}. 
In fact, with the X-ray telescope field-of-view aligned towards the
Moon, this configuration eliminated the $\sim 3.3\times$~ brighter diffuse  
X-ray background radiation (XRB) (see Fig.~4 in Ref.~\cite{moon1}).
ROSAT has also observed  rather intense X-ray emission coming from 
the optically dark side of the Moon (Fig.~2) : its {\it shadow} emits X-rays at a 
level as high as  $1\%$  and  $30\%$  compared to that of the bright side of 
the Moon and  of the  XRB radiation, respectively.
It is interesting to note that all these three components are extracted from
the same X-ray image, and they have a quite similar spectral shape (Fig.~3).

Within the  axion scenario (see section 2.5),  X-rays  should come  (obliquely) 
from the Sun  neighbourhood.
An (isotropic) X-ray emission will mimic scattering of solar X-rays by gas, 
plasma, dust, etc.
\cite{phillips1},
which is a well-known phenomenon in astrophysics. The low-energy 
spectrum of the ubiquitous diffuse X-ray background radiation (XRB)  overlaps 
with the solar one (see Fig.~3), 
while its  intensity is by no means negligible. 
Thus,  reprocessing of solar X-rays by matter in outer solar space and the XRB
must be taken into account, in order to correctly interpret existing data  or
to design a new measurement following the reasoning of this work. 
 
The interaction of the solar wind with the dark surface of the Moon
has been suggested as the source of these lunar X-rays
\cite{moon1}.
Such an explanation should actually be the natural one, but the estimated 
absolute intensity seemed  not to be  completely satisfactory. 
Later, EUV lunar data
\cite{flynn8xxx}
have shown that the decline in albedo from FUV to X-ray
wavelengths was more shallow than expected, and this was considered as a
possible evidence of lunar X-ray fluorescence (not reflection).
Furthermore, Ref.
\cite{moon3}
reaches the tentative conclusion {\it that there was no emission from the
Moon itself}, without excluding the opposite 
\cite{moon1}. 
In addition, a preliminary re-analysis
\cite{moon2}
of this  experimental result reached
the following tentative conclusions 
\footnote{
The purpose was actually to propose a possible
extra component of the soft XRB
\cite{moon2}.
} :
\begin{itemize}
\item[a)] The dark side of the Moon is brighter than expected  by a factor 
of more than  10.

\item[b)] This excess is consistent with the  effect of an X-ray emitting
region around the Earth.
\end{itemize}
\noindent
From ~all these conclusions, it seems that a ~conventional reasoning ~has
difficulties to convisingly explain this (unexpected) lunar X-ray shadow. 
However, these X-ray photons from the {\it direction} 
of the dark side of the Moon (Fig.~3B) fit the solar axion scenario (with the 
axion energy being obviously twice that of  Fig.~3B).
It is interesting to note that the energy of the bulk of these single
photons (Fig.~3B) from the dark Moon direction ($\sim$ 100 to $\sim$ 400~eV)
coincides with that discussed  in Section 2.1 ($\sim$ 50 to $\sim$ 350~eV),
while the observed X-ray spectrum extends up to $\sim 2$ keV;
beyond this energy the response curve of ROSAT diminishes.

In any case, from this lunar observation we derive the best experimental upper 
limit for the radiative decay rate of exotic particle candidates at the  
site of the Earth. This is a useful input  for (future) experimental searches 
of this kind  (see Section 3).

\subsection{The  X-ray background radiation (XRB)}

 The origin of the diffuse XRB radiation accidentally discovered  
 in 1962
\cite{ref1,trumperX}
remains a mystery
\cite{day}
 and is the most enduring problem in  X-ray astrophysics 
\cite{mchardy,suto8xxx}.
From a wealth of data about its galactic component below 1 keV, the
existence of a $\sim 100$~pc extended optically thin local $\sim 10^6$~K
($\sim$ 0.1 keV) hot
plasma has been invented, which fills the `Local Hot Bubble'
(LHB)
\cite{breit} :   
a soft X-ray-emitting region around the Sun, essentially free of neutral gas
\cite{welshxxx},
whose origin is unknown
\cite{breit,egger,sfeir}. 
The existence of a hot gas as a major constituent of the 
interstellar medium follows from the  observed UV resonance absorption lines 
from highly ionized species such as O-VI
\cite{wang},
while recent work
\cite{sfeir}
suggests either a much smaller temperature of the gas, or that it is
hot but also extremely metal deficient.
Unexpectedly, no emission lines from such a hot plasma were detected,
and {\it this result is disturbing}
\cite{bowyer1xxx}.
However, in the suggested axion scenario, a radiatively decaying population of
trapped axions mimics a hot plasma (component), and it remains hidden.

The soft XRB temperature component of $\sim 10^6$~K
\cite{LHB}
is encountered also in various other places in  the Universe.
In the $\sim 0.1$ - 0.8 keV band the XRB intensity exceeds  the extrapolated 
higher energy part
\cite{suto8xxx,kuntz1xxx,barberxxx,miyaji777x,has7777,parmar}; 
a  broad  diffuse soft excess (below $\sim 1$ keV) seems to be established
\cite{KUSHINOXYZ}.
Orbiting X-ray telescopes have found ubiquitous plasmas above 
$\sim 10^6$K, whose heating, however, is poorly understood
\cite{gudelxxx},
while most of the baryons in the Universe are thought to reside in 
intergalactic space at temperatures $\sim 10^{6\pm 1}$K
\cite{voit}.

Observations show that the  XRB is comprised of the integrated contributions 
from a large number of discrete sources (including Clusters of Galaxies)
\cite{parkX,hasinger,hasingerxxx},
with only a small portion remaining unexplained
\cite{phillips1xxx}.
In addition, the recently obtained XRB spectrum
\cite{kuntz1xxx,miyaji777x,phillips1xxx},
with a bump at $\sim 0.7$ keV, coincides in energy with the bulk of the 
spectrum of  solar-type Stars
\cite{gudel3xxx}.
The (same) axion scenario should apply though not only to {\it our} Sun.
In summary, the following observations are in favour of an
axion related scenario as the origin of the (soft) XRB : 

\begin{itemize}
\item[a)] 
The quiet-Sun corona 
\cite{orlando,alexander,phillips,sf} and
the coronae of other Stars
\cite{gudelxxx}
(most late-type Stars are surrounded by hot (above $10^6$K) corona
\cite{fleming7xxx}).
\item[b)]
The Local Hot Bubble
\cite{jelinsky,snowdenX,snowdenZ},
and the galactic diffuse X-ray emission beyond it
\cite{parkX,snowdenX,snowdenZ,sumnerX}.
\item[c)]
Intergalactic matter, which requires a significant amount of non-gravitational 
heating (see below).
\item[d)]
A number of X-ray Clusters  of Galaxies are better modelled 
using a second cooler component of $\sim 10^6$K
\cite{lieuxxx}. 
Also, a soft excess ($\sim$ 0.1 - 0.4 keV) from the central Galaxy M87 
of the Virgo cluster has been established; in order to prevent 
the gas from rapidly  cooling,  
{\it some as-yet-undiscovered heat source} must be at work
\cite{buote,fabianxxx,bohringer,borgani5xxx}.

\end{itemize}

\vskip0.3cm

For the extragalactic component of the XRB above $\sim$ 1 keV
\cite{parmar},
where the bulk of its energy density resides
\cite{vignali777x},
a  widespread optically thin and very hot uniform 
intergalactic medium  (temperature $\sim 10$--40~keV) has been ruled out 
by the lack of distortion of the  Planck spectrum of the 2.7~K Cosmic Microwave
Background radiation (CMB)
\cite{hasinger,mather,henry2,fabian} : 
the Compton scattering with the very hot plasma electrons should generate a 
change in the CMB temperature (Sunyaev-Zel'dovich effect)
\cite{parmar}.
For this XRB radiation, an axion scenario could also  be at work, requiring
however  accordingly hot  places  in the Sky.

\subsection{The soft X-ray excess phenomenon}

Decay photons from axions gravitationally captured in the potential wells of
galaxies, or clusters of galaxies have been searched for in the past as 
{\sf narrow lines} in  microwaves 
\cite{blout}
as well as in the visible
\cite{bershadyx}.
By contrast, the radiative decay  of {\sf massive} axions or axion-like 
particles should result primarily in a diffuse  photon emission  with a 
\underline{broad} energy distribution  reflecting the temperature of their  
place of birth (e.g. the Sun core, or the accreting disk of a massive
black hole, etc.). In addition, if the observed X-rays happen to be in excess
of the conventionally expected ones, we are led to further consider whether
such an observation is in favour of this work
\footnote{
It is not unreasonable to assume that  axions or other axion-like particles 
dominate the potential wells of astrophysical objects like Galaxies, or Clusters
of Galaxies where dark matter was first localized by Zwicky in 1933
\cite{zwicky}.
In order to make a rough estimate of their luminosity, we use for the axion 
lifetime ($\tau$) and the total mass (M) of a group of Galaxies the values
$\tau \approx 10^{17}sec \approx$ 3~Gyears
and M=10$^{13}M_{\odot}\approx 10^{67}$ erg, respectively. Thus, the derived  
luminosity is  L$_x\approx 10^{67}/10^{17}\approx 10^{50}$ erg/s.
For comparison, the observed values  are at  L$^{obs}_x\approx
10^{42}$ erg/s. Therefore, even if the assumed axion abundance and/or their 
decay rate is by several orders of magnitude smaller,  X-rays from such huge
astrophysical agglomerates can still  provide a sensitive signature  for the 
putative unstable particles.
}.
As usually in astrophysics, there is, however, no luck of alternative
explanations. Therefore, we address (throughout this work) the mostly striking 
findings we are aware of, which lack any clear explanation.

We emphasize, among many astrophysical observations of a soft X-ray excess
\cite{excessxxx,romanoxxx}, 
the recent discovery by  ROSAT of the so far brightest  
{\sf diffuse soft X-ray excess} from a cluster of Galaxies 
\cite{bonamentexxx}
\footnote{
Sersic 159-03  with : L$_x \approx 10^{45}$ erg/s below 2 keV, 
redshift z=0.056, 
radial size of the X-ray emitting region $\approx$ 800 kpc; 
the maximum iron abundance (=: metallicity) is 
Z=0.55$\times Z_{\odot}$ in the central part 
\cite{kaastraxxx},
decreasing exponentially outwards ($Z_{\odot}$=solar metallicity). 
The dominant temperature is T$\approx 2.6$ keV;
a cooler component has T$=(0.8\pm 0.2)$ keV and contributes by $\sim 2\%$ in 
X-rays to the total X-ray luminosity.

\noindent
The soft excess phenomenon may be a common occurence in Galaxy Clusters
\cite{MBXYZ}.} :
since all the competing models are confronted with serious problems,
{\it a major effect must be at work in the intergalactic medium that has
hitherto been completely ignored}.
In addition, models of cluster formation, in which the intergalactic gas falls
into the dark matter dominated gravitational well, fail to reproduce basic
observed properties. Similarly, also other authors  
\cite{bialekxxx}
conclude that
{\it there appears to be additional  physics driving Intra Cluster Medium} (ICM).

In the following,  we also mention some at first sight not so obviously 
relevant properties of astrophysical plasmas, which might well be of relevance, 
in particular when they are seen collectively :

\noindent
{\sf a)} The {\sf entropy excess}
\cite{lowensteinxxx,mulchaey},
along with the estimated short cooling time compared with the age of the medium
\cite{lieuxxx},
provided evidence for some kind of non-gravitational heating of the ICM. 
Most viable sources of heating seem to be insufficient in
bringing the ICM to the observed entropy level, i.e., there is an {\it energy
crisis}
\cite{tozzixxxxx}.

\noindent
{\sf b)} The unusually {\sf low metallicity}
\footnote{
For example : a) the metallicity of the nearby Galaxy NGC 1291 
measured recently with CHANDRA is Z=(0.04$\pm$0.02)$\times$Z$_{\odot}$
~\cite{irwinxxx}; 
b) groups of Galaxies have Z=(0.06 - 0.15)$\times$Z$_{\odot}$,
which should be compared with Z=(0.2 - 0.3)$\times$Z$_{\odot}$
found in  clusters
\cite{mulchaey}.
One would naively expect the opposite, since the ratio of stellar mass
to gas mass is higher in groups than in clusters of Galaxies
\cite{mulchaey}. 
}, 
i.e. low elemental abundance
\cite{kaastraxxx,mulchaey,tozzi,xia,read,hxu},
could  also come from an overestimated mass of the 
gas. Note that for a thermal plasma the X-ray luminosity is a measure of the 
gas mass. The radiative decay of captured exotica by the same gravitational well
pretends additional radiating gas, resulting in a lower metallicity
\footnote{Note also that most elliptical and other early type Galaxies possess a
hot diffuse interstellar medium (T$\approx $0.5 - 1 keV). Equally puzzling with
their low metallicity is the observed strong disagreement with the expected
radiation cooling (cooling time $\leq 10^8$ years). In addition, there are no
obvious sources of  gas heating  
\cite{hxu}.
}.
Even though a mixture of gas temperatures could provide an alternative
explanation,
according to a recent review  article 
\cite{mulchaey},
{\it the metallicity of the intragroup medium remains an open issue}
\footnote{
A group is a small cluster of galaxies.
}.

\noindent
{\sf c)} The observed {\sf X-ray Luminosity-to-Gas Temperature relation}  
in clusters of Galaxies is inconsistent with the expected $L_x \sim T^2$
relation, especially for small groups of Galaxies
(e.g., it was found that   $L_x \sim T^8$, or, $L_x \sim T^5$)
\cite{mulchaey}. 
Again, X-ray emission from gravitationally trapped massive axion-like 
particles  can cause ~$-$in principle$-$~ any  discrepancy
\footnote{
The X-ray temperature from clusters of Galaxies measures also the depth of
the gravitational well (given mainly by the dark matter), providing thus a 
crucial link between the physics of the plasma and that of the dark matter 
condensations
\cite{cavaliere6xxx}. 
If the decay or any other process associated with some exotica
gives rise to the emission of X-rays, 
then, this link will be falsified  depending on the intensity of
these  additional photons.

\noindent
For example, for the central $\sim$ 200 kpc of a Cluster of Galaxies a
significant mass discrepancy between X-ray and gravitational lensing methods
seems to exist
\cite{sjxue} :
$\frac{lensing~mass}{X-ray~mass} \approx 2$.
The mass ($M$) enclosed within a radius $R$ is given by the relation
$M(R) \sim R^2\cdot T$.
Thus, from a modified plasma temperature $T$ the derived mass will be wrong
accordingly. Interestingly, the two methods yield consistent result for large
radii.
}.
Note that a significant steepening appears to occur below 
$\sim 1$ keV. 
The maximum deviation appears for groups of Galaxies with 
T$\approx$0.3 keV, and this was considered as an indication for 
non-gravitational heating of the plasma
\cite{mulchaey,fabrizioxxx}
\footnote{Probably even prior to its entry into the dark halos.
};
also, the estimated excess entropy associated with  preheating
corresponds  to a similar temperature 
\cite{ponmanxxx,lloydxxx}.
Furthermore, it seems that the central regions of small Clusters and groups of 
Galaxies  require a larger non-gravitational energy injection of 
$\sim 1$ keV per particle
\cite{borgani4xxx}
\footnote{
Within the axion scenario, this higher value should require more axions, 
i.e. more dark matter at the central region, implying a deeper gravitational 
well, which is not so unreasonable to happen.
}.
Thus, the axion scenario can accommodate the required additional  energy source 
and the X-ray luminosity
\cite{mulchaey}.

\subsection{Gravitationally trapped axions as a source of solar X-rays}

If the solar X-rays observed near the Earth originate from two-photon
decay of axions produced in the Sun core, then the space around the
Sun becomes a source of X-rays. In this case one expects a correlation
between the solar X-ray flux, the axion density and the axion mean
lifetime.

In this scenario the mean axion decay length must be much shorter than
the Sun-Earth distance, because we know that most of the solar X-rays
originate from a region near the solar surface. So, a mean axion
lifetime $\tau _a$ of the order of one minute, or shorter, is needed.
The $a \rightarrow \gamma \gamma$ decay rate is
\cite{raf1990}
\begin{equation}
\tau _a^{-1} = \frac{g_{a\gamma\gamma}^2 m_a^3}{64\pi}
\end{equation}
where $g_{a\gamma \gamma}$ is the $a$-$\gamma$-$\gamma$ coupling constant
and $m_a$ is the axion mass. For masses around 1~keV and a mean lifetime
of 1~minute $g_{a\gamma \gamma}$ is $\sim 1.5 \times 10^{-3} GeV^{-1}$.
With such a value the mean free path for $a\rightarrow \gamma$ conversions 
by inverse Primakoff effect inside the Sun ($aZ \rightarrow \gamma Z$, see
ref.
\cite{raffelt8xxx}) 
is much shorter than the Sun radius and no axions can emerge from the Sun
\cite{dolgovxxx}.

This inconsistency can be avoided by assuming that the main source
of solar X-rays consists of  accumulated long-lived axions, which are 
gravitationally trapped in closed orbits around the Sun. In this scenario
$g_{a\gamma \gamma}$ can be small and, therefore, the axion interaction mean free
path in the Sun becomes  extremely long (see below in this section). 
Axions with lifetimes as long as the present age of the solar system 
($T_s \approx 4.6 ~Gy$),
or longer, are acceptable because sooner or later they all decay in the vicinity
of the Sun by definition. However, it is obvious that in this framework
axions must not have a unique mass value, because in this case the trapped
axions have very low velocities (lower than the escape velocity,
$v_{esc} = 6.175 \times 10^5 m/s$), and they decay to X-rays which
are almost mono-energetic. In addition, Peccei-Quinn axions from the Sun 
are expected to be relativistic, and, they escape from the gravitational field
of a Star like our Sun.

In order to investigate this scenario we have used a simulation program
based on the Kaluza-Klein (KK) axion model described in ref.
\cite{dilella5xxx}. In this model the lightest axion is identified
with the conventional Peccei-Quinn axion and there is an infinity of 
heavier KK excitations separated in mass by 1/R, where R is the
compactification radius (of the order of 1 eV$^{-1}$).

In this model axions are produced by the mechanisms of photon coalescence
($\gamma \gamma \rightarrow a$) or Primakoff effect
($\gamma Z \rightarrow aZ$). 
The total solar axion luminosity $L_{a}$ (namely the rate
of solar energy produced in the form of axions) is given by
\begin{equation}
L_{a} = A \cdot L_{\odot} \cdot \left(\frac {g_{a \gamma \gamma}} 
{10^{-10} GeV^{-1}}\right)^2\cdot
\left(\frac {R} {keV^{-1}}\right)^{\delta}
\end{equation}
where A is a numerical coefficient, $L_{\odot}=3.85 \times 10^{33}$ erg/s
is the standard solar luminosity, and $\delta$ is the number of extra
dimensions.
We use $\delta = 2$ and $R = 10^3 keV^{-1}$
\cite{dilella5xxx}. 
With this choice the numerical values of A and the average energies of the 
produced axions ($\overline{E}_a$) are given in Table 1.  

In our simulation axions with a continuous mass distribution between
0.01 and 20 keV are isotropically generated at different radii inside
the Sun  and are traced through
the Sun by numerical integration of the equations of motion. The radial
dependence of the solar temperature and density are taken from
ref.
\cite{bahcall}. 
At the Sun surface only axions with velocities $v$
below the escape velocity $v_{esc}$ are
considered further. Axions with $v > v_{esc}$ are discarded because
they leave the solar system before decaying. Typical orbits of
trapped axions are shown in Fig.~4.

The fractions of trapped axions for the two production mechanisms
are given in Table 1 for the two production mechanisms. The
trapping fraction is much smaller for axions produced by Primakoff
effect, because in this case the target is at rest and it is more
difficult to produce low momentum axions.
 
Fig.~5 shows the velocity distribution at production for axions
produced by the mechanism of photon coalescence. The trapped axion mass
distributions for the two mechanisms are shown in Fig.~6.

We note that in this model it is assumed that the electric charges are
isolated and the initial state photons are massless.
Both assumptions are not correct because the effective photon mass 
in the Sun core is given by the plasma energy
\cite{raffelt8xxx}, 
which is typically of the order of 300 eV. 
A non-zero photon mass is likely to affect the results of
our simulation especially for gravitationally trapped axions which
are produced with low velocities by definition. A more correct model
should take this effect into account. Our intention here is to limit our
study to a qualitative assessment of this scenario. 

The radial density of the axions gravitationally trapped around the
Sun is reconstructed from the distribution of the parameters
describing the elliptical orbits outside the Sun. In this scenario
the total number of trapped axions is an increasing function of time
\begin{equation}
N_a(t) = R_a \tau _a (1 - e^{-t/\tau _a}) ~,
\end{equation} 
where $R_a$ is the trapped axion production rate under the simplifying
assumption of a steady Sun. Obviously, both $R_a$ and $\tau _a$ depend on
the axion mass $m_a$ and on the coupling constan $g_{a \gamma \gamma}$.
The present axion decay rate is then
\begin{equation}
D_a(T_s) = R_a (1 - e^{-T_s/\tau _a})~~.
\end{equation}

For a given value of $g_{a \gamma \gamma}$ we can predict the present
density of trapped solar axions as a function of the distance from the
Sun using Eqs.~(1), (2) and (3). From this distribution, using Eq.~(4)
we predict the present solar X-ray spectrum on Earth and the apparent solar
X-ray luminosity ($L_x$) by calculating the X-ray flux from axion decay
through a spherical surface of 1 AU radius centred at the Sun. We then
determine the value of $g_{a \gamma \gamma}$ by requiring that our predicted X-ray
luminosity be equal to the experimentally reconstructed one, i.e.
$L_{x}^{2-8keV} \sim 10^{23}$ erg/s (this value corresponds to
the solar X-ray spectrum reconstructed for the  ASCA/SIS  detector at Sun
minimum and  integrated over the energy interval from 2 to 8 keV 
\cite{orlando} (see also below)). 
This procedure gives $g_{a \gamma \gamma} = 9.2\cdot 10^{-14}~GeV^{-1}$.
With such a value of $g_{a\gamma \gamma}$, the axion interaction cross section 
via the Primakoff effect is below $\sim 10^{-54}~cm^2$
\cite{raffelt8xxx}, 
corresponding to a mean free path  much larger than the total trapped 
axion flight path even for the age of the Universe. 
The apparent X-ray luminosities and total axion luminosities obtained from 
this value of $g_{a\gamma \gamma}$ are given in Table 1. 
Fig.~7 shows the trapped axion density as a function of the distance from the 
Sun (for $g_{a \gamma \gamma} = 9.2\cdot 10^{-14}~GeV^{-1}$). 
The predicted X-ray spectrum is displayed in Fig.~8.
We note that in this scenario, and with the value of $g_{a \gamma \gamma}$
given above, the total solar axion luminosity (for both production mechanisms)
is $L_a = 6.1 \times 10^{32}$ erg/s (see Table 1), 
which is $\sim 16\%$ of the standard solar luminosity $L_{\odot}$. 
\begin{table}[H]
\caption{List of parameter values}
\[
\begin{array}{|l|c|c|}  
\hline
&\mathrm{Photon\ coalescence}&\mathrm{Primakoff\ effect}\\
\hline
\multicolumn{1}{|c|}{f_{trap}}&9\times 10^{-8}&5\times 10^{-11}\\
\hline
\L^{2-8keV}_x (erg/s)&1.0 \times 10^{23}&2\times 10^{19}\\
\hline
A&0.067&0.12\\
\hline
\overline{E}_a (erg)&1.1\times 10^{-8}&6.2\times 10^{-9}\\
\hline
L_{a} (erg/s)&2.18\times 10^{32}&3.91\times 10^{32}\\
\hline
\end{array}
\]
\end{table}
\noindent
It is known that  any exotic energy loss process in the Sun is overcompensated
by an increased consumption of nuclear fuel in the core  and results, therefore,
in a temperature increase in the core
\cite{RAFFELTUUU}.
Until recently, this temperature was mainly constrained by helioseismology
\cite{hschlattl},
excluding any exotic energy loss process much in excess of the $0.2L_{\odot}$
level. However, the SNO experiment has very recently measured the total neutrino
flux from $^8B$ decay in the solar core independently of the neutrino flavour
\cite{SNOXXX}. This flux, $\Phi _{\nu} = (6.4\pm 1.7)\times 10^6/cm^2s$, 
provides a more stringent constraint to the temperature of the solar core
($T = (15.74 \pm 0.19$) MK) and, therefore, to the energy loss rate from any 
exotic process, which is now limited to 
$(0.037 \pm 0.063)L_{\odot}$
\cite{hschlattl};
i.e., the value of $L_a=0.16L_{\odot}$ is still allowed within $2\sigma$.
Because of the rather crude assumptions involved in our generic model, we do not
consider this limit as a serious disagreement with the scenario suggested here.
One day, if axions or other particles with similar couplings are discovered and
neutrino data are more accurate, they can provide very important mutual
constraints.

We also note that the axion mean lifetime is much longer than the age of the 
solar system: as an example, for an axion mass of 5 keV we find 
$\tau _a = 1.25 \times 10^{20}$ s $\approx 4000$ Gyears.  
Such a lifetime is actually quite short for conventional axions. 
An axion related solar X-ray luminosity  ~$-$as it is advocated in this 
work$-$~ contains then the whole Sun history since its birth (or even
before?). 
With these values for $L_a$, $\tau _a$ and the mean axion velocity being
$v \approx 0.6~c$ (Fig.~5), we calculate, for comparison, the X-ray luminosity 
within  1 AU around the Sun  coming from escaping axions decaying in flight :
$L_x^{2-8 keV}(escaping~axions) \approx 
\frac{10^3s}{10^{20}s}\times 6.1\cdot 10^{32}~erg/s \approx 10^{16}$~erg/s.  
This makes only $10^{-7}$ of the X-ray luminosity expected from
the decay of the same but trapped solar axions (see Table 1).

An additional comparison is of interest. If an axion
 related luminosity of $0.16 L_{\odot}$ applies to all stars, and the
 axion lifetime is $\sim 400\times$ the age of the Universe, this gives a
 ratio

 \begin{center}
 $\frac{total~photon~energy~from~axions}{total~photon~energy} \approx
 \frac{0.16}{400} \approx 0.4\cdot 10^{-3}$
 \end{center}
 
 \noindent
 From the wide band cosmic energy  density spectrum 
 \cite{HAUSERXYZ}
 the ratio between the 2-10 keV range and the visible band is
 $\sim $2 to 3$\times 10^{-3}$. Resolved  X-ray sources 
 account for $\sim 80\%$  of the cosmic hard X-ray background (XRB)
 \cite{KUSHINOXYZ}. 
 However, only $\sim 40\%$ of the hard X-ray sources (mainly AGN) are luminous
 and the rest have faint or, in some cases, undetectable, optical counterparts
 (see Mushotzky et al. in ref.\cite{hasingerxxx}).
 Then, for the remaining (diffuse) component this ratio becomes 
 $\sim 10^{-3}$, which is consistent with the 
 axion-to-photon energy ratio ($\sim 0.4\cdot 10^{-3}$) of this work.
 Interestingly, our generic axion model along with the assumption that the
 Sun is representative for all stars in the Universe agrees well with
 the cosmic energy density spectrum.

\subsection{Discussion}

In the following we compare the  predictions of the axion scenario
simulation with solar X-ray data, suggesting  also how 
X-ray observatories can  be utilized in this field. 
In so doing, we go a step further from the primary evidence based on the
temperature/density distributions in the solar atmosphere, where the
temperature rises outwards instead of decreasing, and it even exceeds
$10^6$K. We refer to the quiet Sun, i.e. non-flaring Sun, during the 
solar minimum (this does not imply that the 'active' Sun is not of potential 
interest). Other effects of possible relevance are adressed too. Thus :

\vskip0.2cm
\begin{itemize}

\item[1)]  
The simulated gravitational capture rates (see Table 1) for the two axion
production mechanisms are rather reasonable, in particular for the $\gamma
\gamma$-coalescence mode.  The integrated axion luminosities ($L_a$) are
not inconsistent with the solar  luminosity ($L_\odot$)
\cite{hschlattl}.

\item[2)]  
The shape of the reconstructed  solar X-ray spectrum  (see Fig.~9) 
is consistent  with the predicted one from axion decays for energies above
$\sim 1$ keV (Fig.~8). 
The bulk of the observed X-rays is, however, in the sub-keV range. 
From Fig.~9 the solar luminosity in the 2 to 8 keV band
is a factor of $\sim 10^2$ lower than that between  0.5 and 2 keV,
i.e. with $L_x^{0.5-2keV}\sim 10^{25}$ erg/s  at solar minimum
\cite{orlando},
it follows that $L_x^{2-8keV}\sim 10^{23}$ erg/s.

\item[3)]  
In the reconstructed solar X-ray spectrum (Fig.~9), the soft X-ray energy 
"excess" below 1 to 2 keV might be due to a possible electron Bremsstrahlung 
process,  which  can produce axions via the  axion-to-electron coupling  
\cite{raffelt8xxx}.  
Taking into account the results of our simulation with the Primakoff effect,  
it is not unreasonable to assume that KK-axions from this reaction might have 
a restmass distribution around the energy of the corresponding 
PQ-axions of $\sim$ 0.8 keV, i.e. $E_{\gamma} = m_{KK}/2 \approx$ 0.4 keV. 
Their capture rate  should also be  comparable with the Primakoff reaction, 
i.e. much smaller than  the $\gamma \gamma$-coalescence mode (see Table 1).

\item[4)] 
Recently, in the vicinity of the Earth, the XRB radiation in the 2 to 8 keV 
region has been measured by the CHANDRA X-ray observatory 
\cite{cowie222}.
Its flux is
\footnote{The corresponding intensity of the XRB measured with ROSAT 
\cite{moon1} 
is  

$L_x^{0.1-2keV} \approx  0.4\cdot 10^{-11} erg/s\cdot cm^2 \cdot deg^2 $ .}
\begin{equation}
 L_x^{2-8keV} \approx  10^{-11} erg/s\cdot cm^2 \cdot deg^2  ~.
\end{equation}
Let us assume that this flux comes entirely from the decay of trapped axions
at 1 AU from the Sun.
Then, from this flux, part, or even the bulk, of the reconstructed solar X-ray 
luminosity at 2 to 8 keV should be reproduced, but not (much) more, provided 
the X-rays from the Sun are axion-related, and their space distribution is 
given in Fig.~7.  
Then, the isotropic axion decay inside a cone with one square degree 
opening and $\sim 50~R_{\odot}$  height should reproduce the luminosity given by
Eq. (8); the X-ray contribution from further away becomes more and more 
negligible, because of the decreasing axion density.
With simple calculations we arrive at an upper limit of the 
"specific X-ray luminosity" ($S$) in the 2 to 8 keV range of 
\begin{equation}
 S \approx  5 \cdot 10^{-20} erg/s\cdot cm^3  ~.
\end{equation}
This rate at 1 AU from the Sun, multiplied by a factor of $\sim 10^9$
(following Fig.~7), gives the corresponding value near to the solar
surface. The estimated  effective volume around the Sun, where most of the 
trapped particles are  expected to be (see Fig.~7),  is  
~$\sim 3\cdot 10^{33}~cm^3$ .
The integrated X-ray luminosity ($L_x^{2-8keV}$) from the Sun should be
\begin{equation}
 L_{x}^{2-8keV} \approx  3\cdot 10^{33} cm^3 \times 5\cdot 10^{-20} 
 erg/s\cdot cm^3 \times 10^9 \approx  1.5\cdot 10^{23} erg/s ~.
\end{equation}
This order-of-magnitude estimate is somehow surprisingly. Because, 
under the assumption that the bulk of the measured diffuse  2 to 8 keV X-rays 
are  related to trapped solar axion  decays in the vicinity of the Earth according to 
our simulation, we have reproduced  with Eq. (7) the experimentally 
reconstructed  solar  X-ray luminosity in that  energy  range (see Fig.~9 and 
\cite{orlando}).
Such a "coincidence"  over as much as 9 orders of magnidute can still  be 
accidental; by comparison, the $R^{-2}$ law gives instead a change by a factor
of $\sim 10^5$. At present, we only conclude that these two rates, differing by 
several orders of magnitude, are fully consistent with each other, applying the 
same axion scenario over 1 AU, which is a very big distance for the
gravitational trapping scenario of this work. 
A new related experimental result can change the significance of this finding
accordingly.
Interestingly, the same XRB measurements have also provided a clear and so far
inconsistent 
\cite{cowie222, hasin12345}
field-to-field difference of $\sim 40\%$ when pointing to the North or
South celestial pole 
\footnote{The precise  inclination of each partial pointing period relative to 
the connecting line X-ray-Observatory$-$Sun will allow to find out whether 
this difference is possibly relevant or not for this work.}. 

\item[5)] 
The generic simulation of KK-axion production of this work predicts a very
specific density distribution in space (Fig.~7).
This radial profile can be used for test purposes.
To the best of our knowledge, such observations do not exist, at least not on
purpose. 
Orbiting X-ray observatories  with collecting areas up to $\sim$ 2000 cm$^2$,
like ASCA, BeppoSAX, CHANDRA, XMM-Newton, are best 
suited to perform such measurements, covering a large part of the predicted  
radial distribution  of trapped solar axions in the entire energy range  
($\sim 100$ eV to $\sim 10$ keV).  
Furthermore, an X-ray observation with the dark Moon in the telescope
field-of-view  should be repeated, since it allows to unambiguously extract the
presumed local source of X-rays, whatever their origin might be at the end.
In fact, present orbiting X-ray telescopes with very large orbits exclude  
possible secondary  effects (e.g. scattering from the Earth atmosphere), 
which was definitely not the case for the ROSAT dark Moon observation.

The equivalent fiducial volume for such observations can be up to  
$\sim 1000~m^3$ (see below). This is far beyond any feasible terrestrial
$4\pi$ ~ X-ray detector with sub-keV threshold, indicating the potential
of the orbiting X-ray observatories. Thus, without any modification, 
they can operate~ $-$even parasitically$-$ ~as enormously sensitive detectors 
of radiatively decaying particles in  outer space, and probably also 
far beyond (using for shadowing other places than the Moon). 
Moreover, an appropriate re-evaluation of existing runs might allow to
extract relevant physics results.
We suggest that such reanalyses be carried out.

\item[6)] 
It is worth remembering that the solar corona problem 
(see section 2.1) refers mainly to  solar atmospheric temperatures up to 
T$\sim 10^6 K$, or thermal energies  of, say,  $3k$T$\approx  0.3$ keV. 
It is much more difficult to reconcile the emission of X-rays from the quiet 
Sun with energies of $\sim$ 2 to 8 keV with conventional (solar) physics
(Fig.~9). 
In addition, one should also notice the wide and non-thermal shape 
of the reconstructed solar X-ray energy distribution above 
$\sim $ 1 keV; our simulated solar axion spectra (Fig.~8) are rather similar,
while reprocessing effects in the Sun atmosphere smear out the original X-ray 
spectrum, in particular towards lower energies (see Section 2.1~~3)).

\end{itemize}
\vskip0.2cm
\noindent
Since the reconstructed solar X-ray spectrum, in particular below 
$\sim 1$ keV, can not be reproduced completely, we mention some
other possible effects, which can be involved:

\noindent
{\bf a)}~ Reprocessing of the absorbed higher energetic X-rays (see Section
2.1~~3)) can explain, for example, only $\sim 1\%$ of the "excess" intensity at
$\sim 0.5$ keV (note $L_x^{0.5-2keV}\approx 100\times L_x^{2-8keV}$).

\noindent
{\bf b)}~ The inner solar plasma energy is distributed around
 $\hbar\omega_{pl} \approx 300$ eV. 
Even though  quantitative  estimates do not exist, plasma effects 
could have an impact on the reactions occuring inside the Sun, in particular
with photon energies near $\hbar \omega_{pl}$, 
improving possibly the production rates and/or the capture efficiency. 

\noindent
{\bf c)}~ Only in the  vicinity of the Sun (or other similar places in the Sky)
the mean spacing  between the trapped exotic particles can become 
comparable  with their  de Broglie wavelength ($\lambda$) after some
accumulation time. 
If the bulk of  these hypothetical particles is identical bosons
(the KK-states are actually not), 
a phase transition to a  Bose-Einstein Condensate (BEC)  could take place.
With a BEC we know that a wide range of very unusual phenomena occur
\cite{dalfovoxxxxx,andersonxxxxx,reinhardtxxx,collinsZ}, 
like collapse, explosion, vortices, matter-wave amplification 
("bosonic stimulation")
\footnote{
In a large density and under appropriate conditions, stimulated decays of axions
to 2 photons ("axionic laser") of opposite momentum has been considered
\cite{axionclumps}.
}
, outward going shells, collimated jets, "miniature supernova", etc.  .
Such phenomena take place also in astrophysics and some of them  even on our Sun.
Even though these considerations are speculative,  it has been argued  already 
that axions can occur in the early Universe in the form of a Bose condensate
\cite{relicxxxxx},
while recently  it has been suggested 
that the dark matter in the Universe could exist  in  the form of BEC 
\cite{becZ}.

\noindent
{\bf d)}~ Cosmic axion Stars with strong interaction effects
with magnetized media, etc., have been considered
\cite{axionclumps,iwazakixxxxx}.
This shows that high density axion clumps can interact  efficiently with 
magnetic  fields, which reach $\sim $kGauss values only near to the solar 
surface. In this context, we mention the observed correlation 
\cite{pevtsovyyy,fisherxxx}
between solar X-ray 
luminosity and magnetic fields in active regions, which has been  
measured with the Yohkoh spacecraft.

\vskip0.2cm
The last two cases (c) and d)) are of potential relevance for the soft X-rays
from the quiet Sun. In fact, if we apply the same reasoning  as for the CHANDRA
X-ray data also to the X-rays measured by the ROSAT PSPC detector, 
the reconstructed X-ray intensity near the solar surface is 
$\sim 10^{-3}$ of the  reconstructed  
X-ray spectrum  of the quiet Sun in the 0.1 to 3 keV band 
($L_x^{0.1-3keV}\approx 10^{27}$ erg/s
\cite{orlando}).
However, if we use the dark Moon data, then the reconstructed part becomes 
$\sim 3\cdot 10^{-2}$, since the considered cone height is 
$\approx 0.5\times R_{\odot}$ instead of 50$\times R_{\odot}$, and, the flux is
smaller by factor $\sim  3.3$
\cite{moon1}.
As long as the reconstructed surface X-ray luminosity is below the observed one 
\cite{orlando},
there is at least no contradiction, even though the reconstructed $\sim
3\%$  is not an unreasonable  value. However, if there is an "excess" in the 
reconstructed solar X-ray spectrum at low energies (Fig.~9), 
this could be due to local  effects taking place only in the  near the Sun 
space, e.g.  cases  c) and/or  d)  could be at work. Concerning  density 
related effects, it is worth noticing that the
mean de Broglie wavelength ($\lambda$) of the captured low-mass axions is by a 
factor of $\sim 10$ larger than the one at the higher masses, since their 
velocity spectrum is quite similar. Since the  relevant number of overlapping 
particles is that within a volume of  $\lambda ^3$, which increases with  
decreasing mass, it is reasonable to expect that density related effects might 
well appear pronounced at lower axion momentum and only near to the surface of
the Sun. 
A contribution from such local effects on the Sun is obviously decoupled from 
that at a remote place; then,  the solar surface X-ray intensity, recostructed
from that at 1 AU, becomes underestimated, and this might happen with the 
"excess" solar soft X-rays in Fig.~9.

\vskip0.2cm

Of course, we consider our model as a suggestion for a possible underlying 
mechanism in the solar atmosphere based on solar axion production rather than a 
quantitative  description of it.
We reiterate that {\sf a detailed analysis of these and/or other 
processes is necessary, in order to provide a complete theoretical
treatment}, which is beyond the scope of this paper.
Therefore, this work might be an opportunity for theory.

\section{DIRECT AXION-DECAY DETECTION}

In the following, we address a few configurations in orbit and on Earth, 
which seem to be the most appropriate ones to directly search for 
axion-like particles.

\begin{itemize}
\item[a)] {\bf An orbiting X-ray telescope} with the Sun being outside its FOV
\footnote{
X-ray telescopes avoid having the Sun in their FOV, 
because  of the $\sim 7\cdot 10^9$ X-rays/cm$^2\cdot$s$\cdot
(\sim $~60-400 eV) arriving at the site of the Earth
\cite{krasno}. 
For example, ROSAT's FOV was pointing at $101^\circ$ away from the Sun.
}
could operate as a solar axion antenna . The dark Moon observation
\cite{moon1}
should be repeated;  during a full lunar eclipse, this might be more 
interesting following (X-ray) background considerations. 
Moreover, pointing an orbiting detector towards the 
{\it dark Earth} while it is in Earth's `night' during each orbit around the 
Earth seems to be a very similar and attractive configuration repeating several 
times per day. A wide detector FOV implies a better signal-to-noise
ratio, or at least a higher sensitivity to detect the radiative decay of 
any (solar) exotica.

We compare a wide-aperture  X-ray detector orbiting at an altitude of 500~km 
and pointing towards the dark Earth with the dark Moon  measurement by ROSAT 
covering a  $\sim 0.5^\circ$  narrow cone.
The efficiency to detect the decay photons is by a factor 
\footnote{
We consider a detector surface
\cite{parmar3}
$A = 200$~cm$^2$ with an opening angle of
$\sim 50^\circ$ orbiting at $\sim 500$~km. Assuming isotropic axion decay,
the effective fiducial volume covered by the detector FOV is equal to
$\int \frac{2}{4\pi r^2}\cdot A \cdot \cos\theta {\rm d}x{\rm d}y{\rm d}z =
\frac{A}{2\pi}\int_{0}^{2\pi}{\rm d}\phi \int_{0}^{\rm 500~km}r^2\cdot 
\frac{1}{r^2}{\rm d}r \cdot
\int_{\cos 25^\circ}^{\cos 0^\circ}\cos\theta\cdot {\rm d}(\cos\theta)$, 
which is $\sim 900$~m$^3$ for a dark Earth configuration
and $\sim 74$~m$^3$  for the dark Moon measurement by ROSAT
\cite{moon1}.
$\theta$ is the angle from the normal incidence on the detector surface and the
factor of 2 comes from the two photons per axion decay. 
}~~
$\sim 900$~m$^3/74$~m$^3=12$ in favour of the assumed dark Earth configuration.
So far, to the best of our knowledge, the dark Earth has been observed with
a narrow detector FOV
\cite{maki},
implying instead a  suppression factor of  $\sim 10^{-3}$ compared with the dark
Moon observation, i.e. hopeless to have observed some signal in the past even 
unintentionally.
Thus, the observed dark lunar emission rate of $\sim 0.15$/s
($E_{\gamma}\leq 2$~keV) by ROSAT
\cite{moon1}
translates  into a rate $R$ for the dark Earth configuration of
\begin{equation}
R \approx 0.15\times 12 \approx 1.8/{\rm s},
\end{equation}
assuming a 200 cm$^2$ orbiting X-ray detector at 500~km with $\sim 50^\circ$
opening angle, which implies a  $\sim 900$~m$^3$ {\it effective} fiducial 
volume within its FOV.
This is actually a rather strong signal.
From the dark Moon observation, the obtained model-independent 
maximum axion decay rate (below $\sim 2$ keV and for non-relativistic
velocities) ${\Large\rm  X}({\bf r})$ at 1 AU from the Sun is 
\begin{equation}
{\Large\rm  X}(|{\bf r}|=1~{\rm AU}) \approx 0.15/s\cdot 74~m^3 
\approx  2.3\cdot 10^{-9}~{\rm axion~decays/s\cdot cm}^3.  
\end{equation}
We note that even if the solar axion scenario is not behind this
particular dark Moon  X-ray observation, the rate given by Eq. (9) remains
valid as an  upper limit for future searches for 
radiative decays of exotic particles in this energy range and in the
vicinity of the Earth. Furthermore, following Eq. (6), the corresponding 
upper limit in the $\sim$ 2 to $\sim$ 8 keV range becomes smaller than that 
given by Eq. (9) by a factor of $\sim$ 100.

\item[b)] {\bf An X-ray detector} with  $\sim 4\pi$ acceptance 
operating on Earth (or, better,  underground) seems to be the most adequate 
direct experimental approach, since it allows reconstruction of axion decays 
inside its fiducial volume by observing  $\gamma \gamma $-coincidences.
Such a detector is actually blind to any direct solar X-rays.
Again, assuming the axion scenario to explain the measured low-energy spectrum 
from the direction of the dark Moon
\cite{moon1}
(Fig.~3B),
and taking into account the rate  derived above (see Eq.~(9)),  the
expected coincidence rate R$_{\gamma \gamma}$ should be measurable for
a rather modest ($20\times 20\times 20$~cm$^3$) fiducial volume:
\begin{equation}
R_{\gamma \gamma} \approx 1.6~{\rm coincidences/d}\cdot (20~{\rm cm})^3 \approx
 200~{\rm coincidences/d\cdot m}^3 .
\end{equation}
We are not aware of  any experimental search of this type in the past.
Because of the widely accepted extremely long lifetime of
the `standard' axions, such a measurement was  meaningless until recently
\cite{dienes,dilella5xxx}.

\underline{Background:} 
Uncorrelated photons  are distributed uniformly over the
fiducial volume while the two photons from axion decay will  convert at close
distance from each other. 
To get an order of magnitude estimate of the background rate from uncorrelated
two-prong events, we take the integral single-prong event rate
($R_{1\rm prong}$) as
measured on the surface in a 1 keV window at 1 keV \cite{giomatarisx} 
using a Micromegas chamber \cite{giomataris}
of dimensions $15 \times 15 \times 0.3$~cm$^3$ :
\begin{equation}
R_{\rm 1prong} \approx 1~{\rm event/s}\;.
\end{equation}
At these energies, practically all photons entering the chamber sensitive
volume interact in the gas, so this rate can be used to obtain the background 
rate ($R_b$) per cm$^3$ inside the chamber for the same energy interval:
\begin{equation}
R_b = R_{\rm 1prong}/\rm (15\times 15\times 0.3)~cm^3 \approx 
0.015/s\cdot cm^3\;.
\end{equation}
If the mean photon absorption length is chosen to be $\sim$~0.3~cm, the
$2\gamma$ signal events will occur within a small cell of volume 
$\Delta x \Delta y \Delta z \sim 1$~cm$^3$. In a Time Projection Chamber
(TPC)  $\Delta x$ and $\Delta y$ are measured directly by
orthogonal electrodes or pads, while $\Delta z$ can be derived
from the time interval between the two signals. In a detector with a sensitive 
volume of 1~m$^3$ there are $10^6$ cells of volume 1~cm$^3$, hence the rate of 
two-prong accidental  coincidences in such a detector is
\begin{equation}
R_{\rm 2prong} =  10^6\cdot (R_b)^2\cdot \Delta t \approx 4.5\cdot 10^{-5}
~{\rm events/s\cdot m^3  \approx 4~events/d\cdot m^3}\;,
\end{equation}
where $\Delta t = 0.2~\mu$s is the drift time over 1 cm, assuming a standard
drift velocity of 5 cm/$\mu$s. Of course, this rate can be reduced considerably
in an underground laboratory. 
We note that photons of $\sim$ keV energies entering the chamber will 
predominantly interact at small distances from the chamber walls. 
Thus the background rate given by Eq.~(13) can be further reduced by requiring 
that the events occur in a fiducial volume at some distance from the walls. 
In addition, for non-relativistic axions, the  equal energies of the
two $\gamma $'s will provide  further background rejection. 
We note that this background varies as the third power of the photon absorption
length which  actually  defines the cell size.
It does not seem unrealistic, therefore, to reach an {experimentally
controllable} background  from two-prong events at a level well below or
comparable with the  expected axion signal at $\sim$ 1 keV.

In order to perform such a measurement, the main detector requirements are 
a)  energy threshold as low as possible, e.g.  $\sim$~100~eV,
in order not to miss a low-energy axion signal favoured actually by the dark
Moon measurement; and
{b)} an adequate space {and} energy resolution, in order to distinguish the 
2$\gamma$ events from background, allowing also to implement  constraints 
from the $a\rightarrow \gamma \gamma $ decay kinematics.
Our preference is to photons in the sub-keV range, or even
below $\sim 400$ eV (see {Fig.~3} and {Fig.~9} of this work,
the high statistics pulse-height spectra of the soft XRB in Fig.~4 of
Refs.~\cite{moon3,pfeffer},
the derived solar(-like) X-ray spectra
(Fig. 5 in
\cite{gudel3xxx}), and section 4). 
However,  X-rays above $\sim$ 1 keV might be not less promising. 
Therefore, a low-density, low-Z ~X-ray detector should be used.
A TPC working at low pressure and/or with low-Z gas, e.g. He,
appears to be a promising detector for this purpose.

\end{itemize}

\section{WHERE ELSE?}

We mention below in short some cases, where axion-like particles
{\it could} be involved, and therefore, they should be followed further.

\begin{itemize}

\item[1)]   a) The night-time ionization in terrestrial or celestial atmospheres :
the measured ionization of the Earth ionosphere at night is larger
than predicted 
\cite{judith}, 
requiring an extraterrestrial source of photons in the UV band. 
b) The ionization and heating of (nearby) interstellar medium
\cite{slavin3xxx,reynolds3xxx,viegasxxx};
the spectrum of the stellar EUV sources is too soft to explain the
observed overionization of Helium with respect to Hydrogen in the Local Cloud
\cite{vallerga3xxx,raud3xxx}, 
which is a mystery
\cite{jenkinsxxx}.

\item[2)]
 Clusters of Galaxies emit in the $\sim 0.1$ - 0.4 keV band substantially in
 excess of that expected from  a hot intra cluster medium at a temperature of a
 few keV
 \cite{bonamente5xxx}. 
 E.g., the intense diffuse excess emission in the EUV in the Coma and Virgo 
 Clusters of Galaxies
 \cite{berghofer1xxx,fabianX}
 is of unknown origin
 \cite{bowyer1xxx};
 the spatial distribution of the EUV flux in each of these Clusters was
 inconsistent with that of a gravitationally bound gas
 \cite{bowyer1xxx}.

\item[3)] 
 The radiative decay of the escaping exotica outside the Sun or other places in
 the Sky resembles  the  scattering of (solar) X-rays off  electrons or 
 dust particles in near space (see Fig.~5.1 in Ref.~\cite{phillips1}).  
 The same reasoning can apply to similar configurations in remote interstellar
 space
 \cite{rolf}
 and to extended X-ray sources
 \cite{soltan}.  
 However,  the recent observations of an X-ray halo with the Chandra
 X-ray telescope provided unexpected results, since they do not fit the
 conventional dust scattering model
 \cite{rksmith}.
 The axion scenario of this work might be the only alternative explanation.

 \item[4)] Underground experiments with threshold in the 
 sub-keV range could provide a direct detection of the putative particles.
 Taken into account the ROSAT efficiency and narrow energy 
 bandwidth, the estimated rate from Eq.~(9) is  $\sim$ 0.1 event/kg/day/keV
 below $\sim 1$ keV.

\item[5)]  Seyfert  Galaxies ($NLS1$) show  a giant soft X-ray excess below 
$\sim 1$ keV
\cite{romanoxxx,boller,poundsxxx}, 
with an X-ray luminosity of 
$\sim 10^{44}$ ergs/s $\approx 10^{17}$$\times$L$_{\odot}^x$ 
and strong variability on time scales of $\sim 1$ day 
\cite{turner2xxx}.
A direct observation of the source size is not yet available, but a simple
blackbody emission is ruled out on the basis of the inferred remarkably small
size of the emission region, exceeding the Eddington limit
\cite{turner2xxx}.

\item[6)] The heating of the intergalactic medium, as its origin is not yet clear
\cite{IGM}.
Also, the warm/hot intergalactic medium between Clusters of Galaxies, which has
not been detected so far neither in emission nor in absorption, it makes
probably  a significant fraction of the baryons in the local Universe 
(z$\leq $1-2)
\cite{phillips1xxx}.

\item[7)] The large-scale diffuse X-ray emission (T$\approx 6$ keV) from the 
centre of our own Galaxy is of as yet unknown origin
\cite{warwickddd}, 
with the giant molecular  cloud Sgr B2  being the strongest diffuse X-ray 
source in this region
\cite{koyamaxxx}~
\footnote{If a diffused plasma distribution around the Galactic Centre is indeed
the source of the  emitted X-rays with a flux 
$\approx 1.1\cdot 10^{-10} ergs/s\cdot cm^2\cdot deg^2$, 
then its temperature and energy density are both too high for the plasma to be
gravitationally bound around the Galactic Centre
\cite{koyamaxxx}. 
In fact, the point sources account for only $\sim 10\%$ of the total flux in
the 2 - 10 keV energy range, indicating strong diffuse X-ray emission
(see Ebisawa et al., in ref.\cite{koyamaxxx}).
}.
Also, the universal X-ray emission of AGNs along with the observed evolution
with cosmic time is of potential interest
\cite{miyaji777x,hornsch777x,fischer777x,h777x,hh777x,lafranca777x};
the exact mechanism which produces the X-rays is not known
\cite{rfmxxx}.
Afterall, massive black holes have the strongest gravitational effects and
make $\sim 0.6\%$ of the bulge mass of a Galaxy
\cite{hh777x}. 

\item[8)] Solar "micro-events"
\cite{benz9xxx},
which seem to be present at all times and also in coronal holes; these are short
emission enhancements (= flares) of $\sim 10^{25\pm 2}$ ergs in quiet regions, 
or, localized brightenings and explosive events. Their msec variability
challenges the understanding of the coronal plasma
\cite{apr},
but also the down-flows along with
the redshifted lines, which are seen in the transition region
\cite{landi9zzz,wikstol9zzz}.

\item[9)]  Solar/celestial jets
\cite{wilsonxxxxx},
including the puzzling X-ray tails behind Pulsars (or even Galaxies?)
\cite{tailxxx}.

\end{itemize}

\section{Conclusion}

A missing  explanation of an astrophysical observation is actually 
suggestive to search for an exotic approach. The framework of the celebrated 
dark-matter physics world is a source of possible exotic  solutions.
In the cases considered  here,  a same axion-like scenario consistently 
explains the usually different alternative  conventional solutions 
(if they exist at all).
This  scenario  is not supposed to abandon globally previous models, 
which describe actually related findings;
it can  be rather complementary, providing a so far missing 
physics input.

A temperature (component) of a few  10$^6$K ($\sim 0.3$ keV ), 
which  appears in  so diverse astrophysical  places, 
such as from the solar corona to Clusters of Galaxies and  probably beyond, 
is associated with several  unexpected significant observations.
In order to  explain this in a combined way, we reach the conclusion that 
some new particles~  $-$we use massive axions 
as a generic example$-$  ~must be involved in processes occurring inside { and}  
outside Stars. 
For example, relatively short-lived massive axions  have  been considered in 
theories with extra dimensions
\cite{dienes,dilella5xxx};
the  two-photon decay mode remains dominant providing  theoretical support
to  our purely  observationally/astrophysically motivated claim of celestial 
axion-like signatures in the $\sim$ keV range. 
In our favoured scenario, axion-like particles escape from their place of birth,
e.g., from the interior of the Sun (or that of other Stars in the Sky), 
get gravitationally trapped and  decay in outer space.

Alternatively, a more or less isotropic radiative decay of other hypothetical 
particles, e.g. massive neutrinos, could in principle also explain the 
astrophysical observations  considered here. 
Only laboratory  experiments  could clarify this issue.
We  give a (theoretically) unbiased  parameter space how and where to directly  
search for such exotica.
Fortunately, the two-photon decay mode allows to have a very high detection 
sensitivity inside a large TPC, because of the much suppressed uncorrelated 
two-prong background  events within a small distance, narrow time and same 
energy.
High-performance low-threshold detectors developed primarily for
high-energy physics experiments can  also be utilized for this kind of 
astro-particle physics.

Following the reasoning for the suggested axion-like scenario,
we should also notice that the previous failed searches
\cite{raffelt333,carlson9xxx}
for axions converted to X-rays inside the external solar/stellar magnetic fields 
do not contradict this work. 
At first, this missing signal can be due to an accordingly small coupling 
strength and/or the required very small axion restmass. Independent on this actually unknown value, the "conventional"
Primakoff effect should result to radially  outwards emitted X-rays,  
excluding  a self-irradiation of the Sun, which we consider as the cornerstone 
of the reasoning of this work. 
The tentative solar Kaluza-Klein  model provides reasonable gravitational 
capture rates,  but  it fails to  completely explain the low energy part of the
reconstructed solar X-ray spectrum.
However, other possible sources like the solar Bremsstrahlung-axions seem 
to have reasonably  low energy, but with a smaller  gravitational capture rate.
For  astrophysical standards, the encountered  discrepancies are actually not
particularly large.

Finally, the estimated axion density due to gravitational trapping  by the Sun
can exceed a critical value, which is necessary for the appearance of a 
BEC,  with unforeseen implications. 
The continuous dynamic coronal phenomen  might be a manifestation
of such processes.
Thus, the predicted particles orbiting around the Sun can become an invaluable
clue to physics beyond the standard (solar) model, explaining first of all the 
as yet mysterious properties of the chromosphere and the corona.
The gravitational trapping of massive particles emitted by the Star
itself provides a  mechanism for the appearance of boson  clumps around a 
Star.

{\sf In conclusion,} the strongest and rather direct evidence in favour of the 
axion scenario  comes from the sofar unexplained solar corona 
related  observations like its  heating mechanism, its narrow
interface to the chromosphere, the chromosphere itself, together with the 
striking similarities of the temperature/density profiles with the 
Sun-irradiated Earth atmosphere. 
The recently experimentally reconstructed solar X-ray spectra combined with 
X-ray measurements from orbiting observatories are the first potential direct 
signatures for this work. They show how these orbiting instruments can directly 
search for decaying particles in (near) outer space, providing also the 
expected maximum rate in future investigations of this kind.
The relevant new parameter introduced by this work is the as yet disregarded
elongation angle of the X-ray Telescope relative to the Sun.
The other astrophysical observations we have addressed, in particular when they 
are seen combined,  provide an additional  piece of evidence, probably of not 
minor importance at the end.

\vskip2.0cm

\section*{Acknowledgements}
 
Constructive criticisms by  Alexander  Dolgov  have stimulated us to go deeper
into the axion scenario. We wish to thank Apostolos Pilaftsis for  
stimulating discussions, and Jim Rich for his comments and suggestions that we
have taken into account in this paper. We also acknowledge the 
help by Thomas Papaevangelou. We thank
Joachim Tr\"umper for the permission to use Figures 2 and 3 from
Ref.~\cite{moon1}. We are
grateful to Judith Lean for indicating to us the references about
the Earth ionosphere and for allowing us to use Fig. 1 from
Ref.~\cite{lean}. We also wish to thank Giovanni Peres for allowing us to use 
Figure 9 of this work and Salvatore Orlando for providing us with 
the original numerical values.
We also thank Tullio Basaglia, Gilda Leoni, Marie-Jeanne Servettaz and 
Jens Vigen from  the CERN library for the great help provided in finding 
many not so easily accessible articles, and some of them even twice.

\newpage

\begin{figure}[H]
\centerline{\includegraphics[width=10cm]{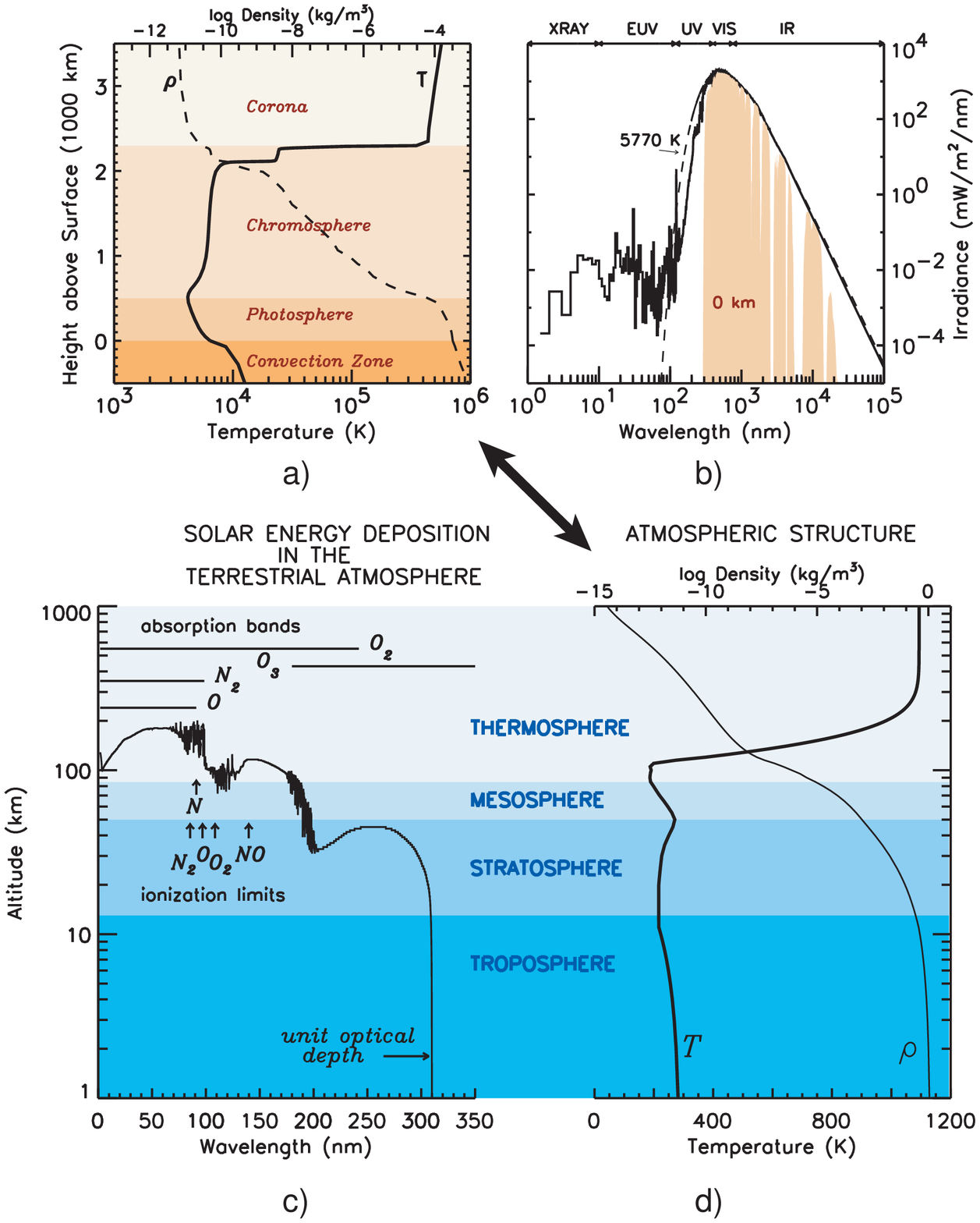}}
\caption{\small ({\bf a})~The mean temperature ($T$) and density ($\rho$) 
 profiles for the solar atmosphere; 
 ({\bf b})~solar irradiance spectrum (the dashed line is the Planck shape for a
 temperature $T=5770$~K); 
 ({\bf c})~the altitude at which
 the Earth atmosphere attenuates the incident solar radiation by a factor
 1/$e$;
 ({\bf d})~temperature ($T$)  and density ($\rho$) as functions of height in the 
 Earth atmosphere. (These figures are taken from Ref.~\cite{lean}.)}
\end{figure}

\begin{figure}[H]
\centerline{\includegraphics[width=10cm]{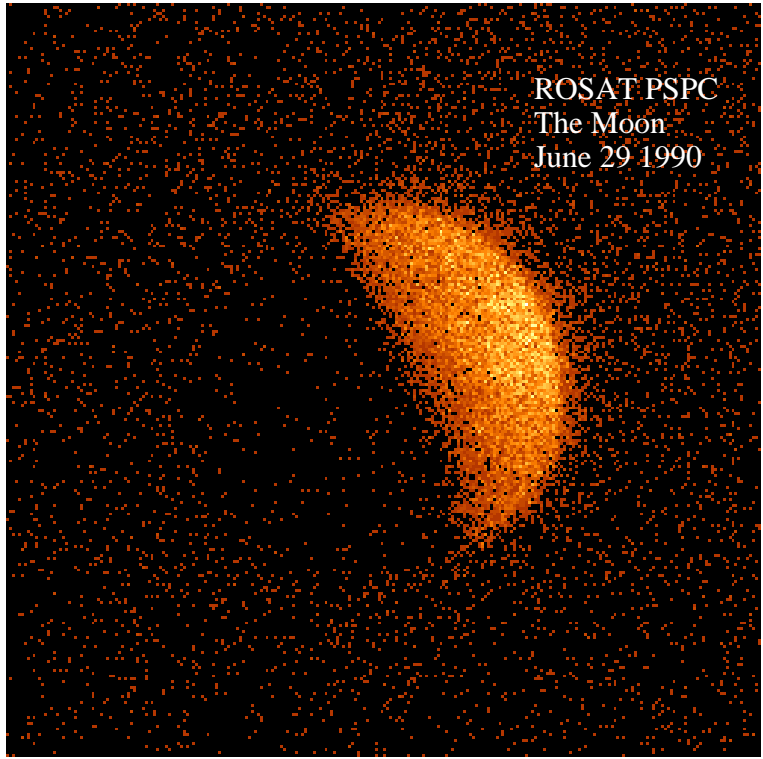}}
\caption{\small The X-ray photon image of the Moon as measured by ROSAT.
 The sunlit portion of the Moon is visible, as well as an X-ray shadow in the 
 diffuse XRB radiation cast by the dark side of the Moon. Grey pixels denote 
 one or two events, except in the brightest part of the crescent, 
 corresponding to three or more counts. }
\end{figure}


\begin{figure}[H]
\centerline{\includegraphics[width=10cm]{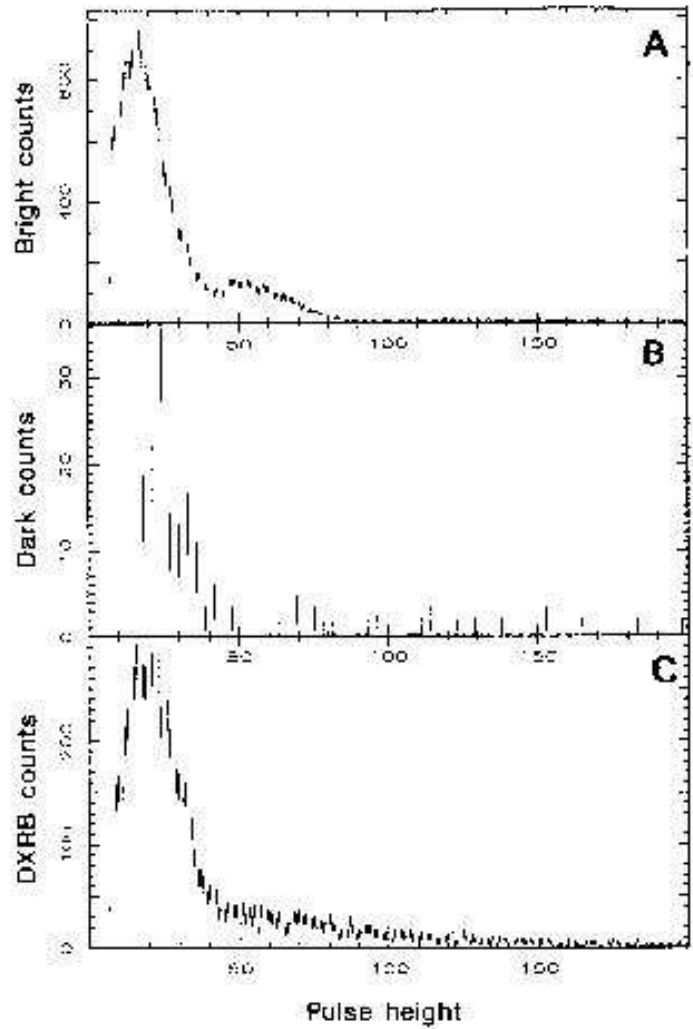}}
\caption{\small Measured raw pulse-height spectra from the observation in 
 Fig.~2 (channel number multiplied by 10 gives the energy in eV, and, the
 effective area of the detector system peaks at the same value of 
 $\sim 200$ cm$^2$ around 250 eV and 1 keV):
 ({\bf A}) X-rays from the sunlit side of the Moon;
 ({\bf B}) X-rays from the dark side of the Moon; 
 ({\bf C}) cosmic X-ray background radiation.  
 (Figures 2 and 3 are taken from Ref.~\cite{moon1}).}
\end{figure}

\begin{figure}[H]
\centerline{\includegraphics[width=10cm]{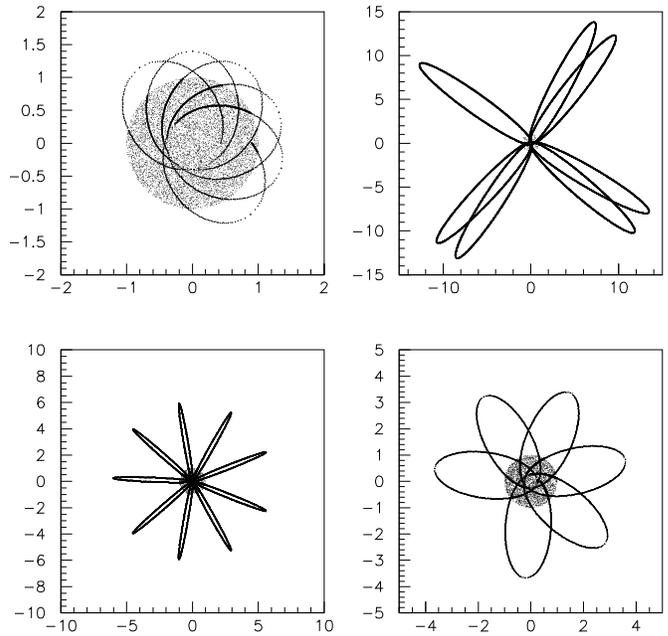}}
\caption[]{Typical orbits of solar axions gravitationally trapped around  
the Sun. The two orthogonal coordinates are given in solar radii. The
shadowed region in the center of each figure outlines the solar disk.
Only the first few revolutions are traced.}
\end{figure}

\begin{figure}[H]
\centerline{\includegraphics[width=10cm]{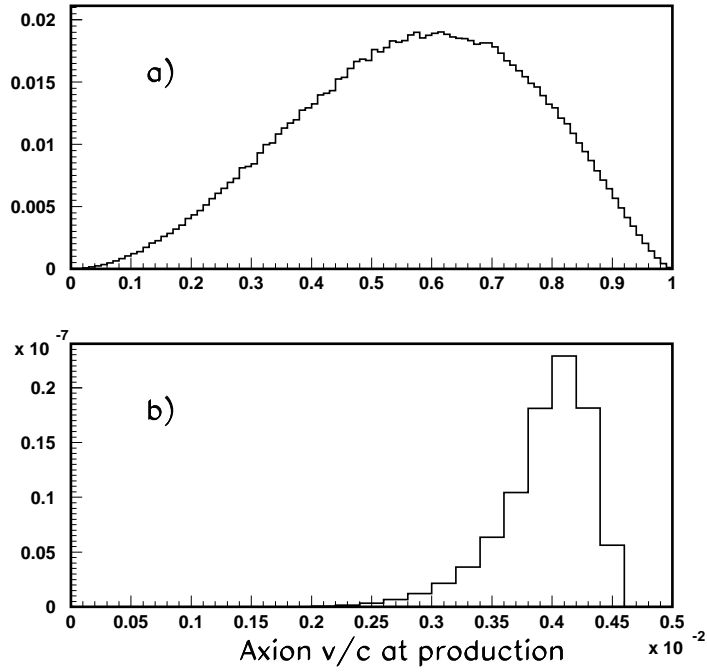}}
\caption[]{Velocity distribution at production for solar axions from
photon ~coalescence : a) all axions (distribution normalised to unity);
b) gravitationally trapped axions (distribution normalised to $f_{trap}$).}
\end{figure}

\begin{figure}[H]
\centerline{\includegraphics[width=10cm]{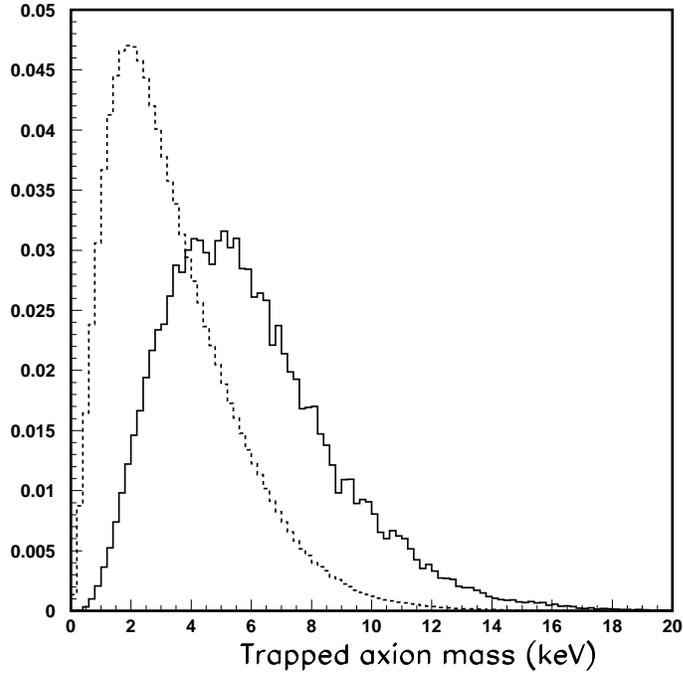}}
\caption[]{Mass distribution of gravitationally trapped solar axions produced by
photon coalescence (full line) and by Primakoff effect (dotted line).
The two curves are normalized to unit area.}
\end{figure}

\begin{figure}[H]
\centerline{\includegraphics[width=10cm]{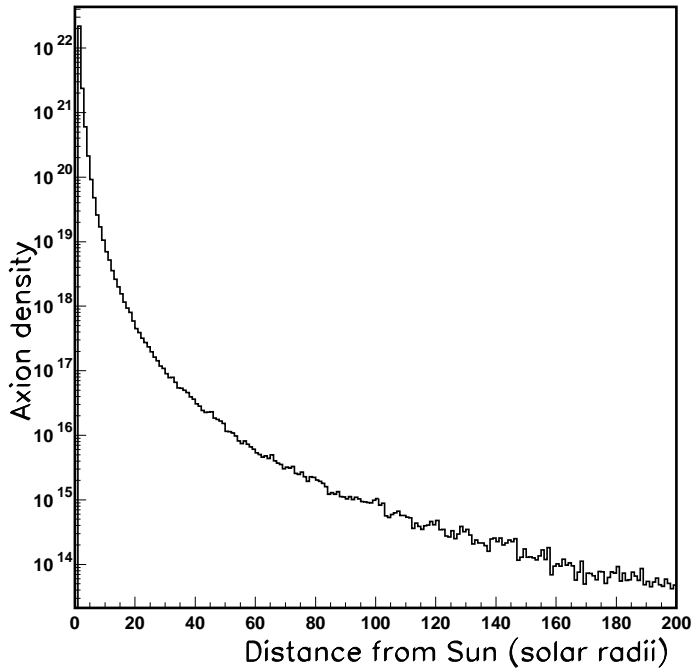}}
\caption[]{Present density (axions per $m^3$) of gravitationally trapped
axions in the region around the Sun, as a function of the distance from
the Sun centre.}
\end{figure}

\begin{figure}[H]
\centerline{\includegraphics[width=10cm]{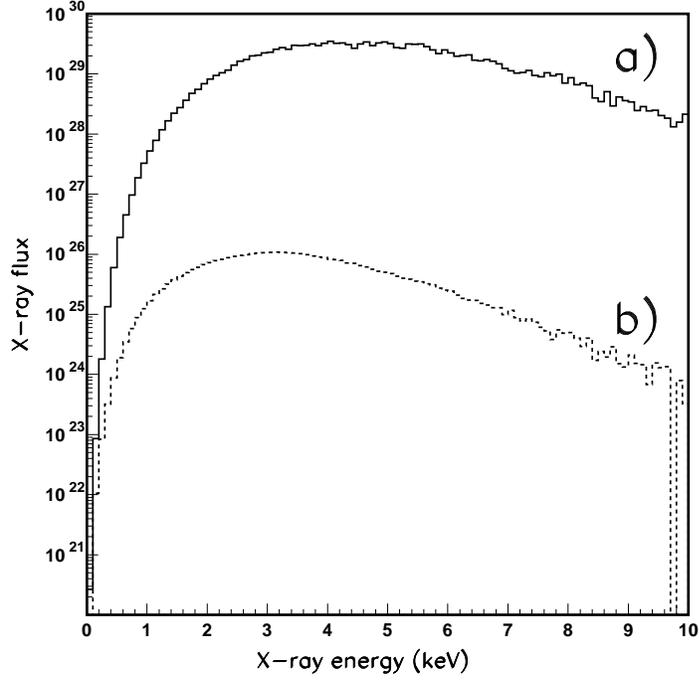}}
\caption[]{Expected X-ray energy spectrum from the decay of gravitationally
trapped  solar axions produced by a) photon coalescence and b) Primakoff
effect. The two curves are normalized to the predictions of our
simulations for the present age of the solar system.}
\end{figure}

\begin{figure}[H]
\centerline{\includegraphics[width=10cm]{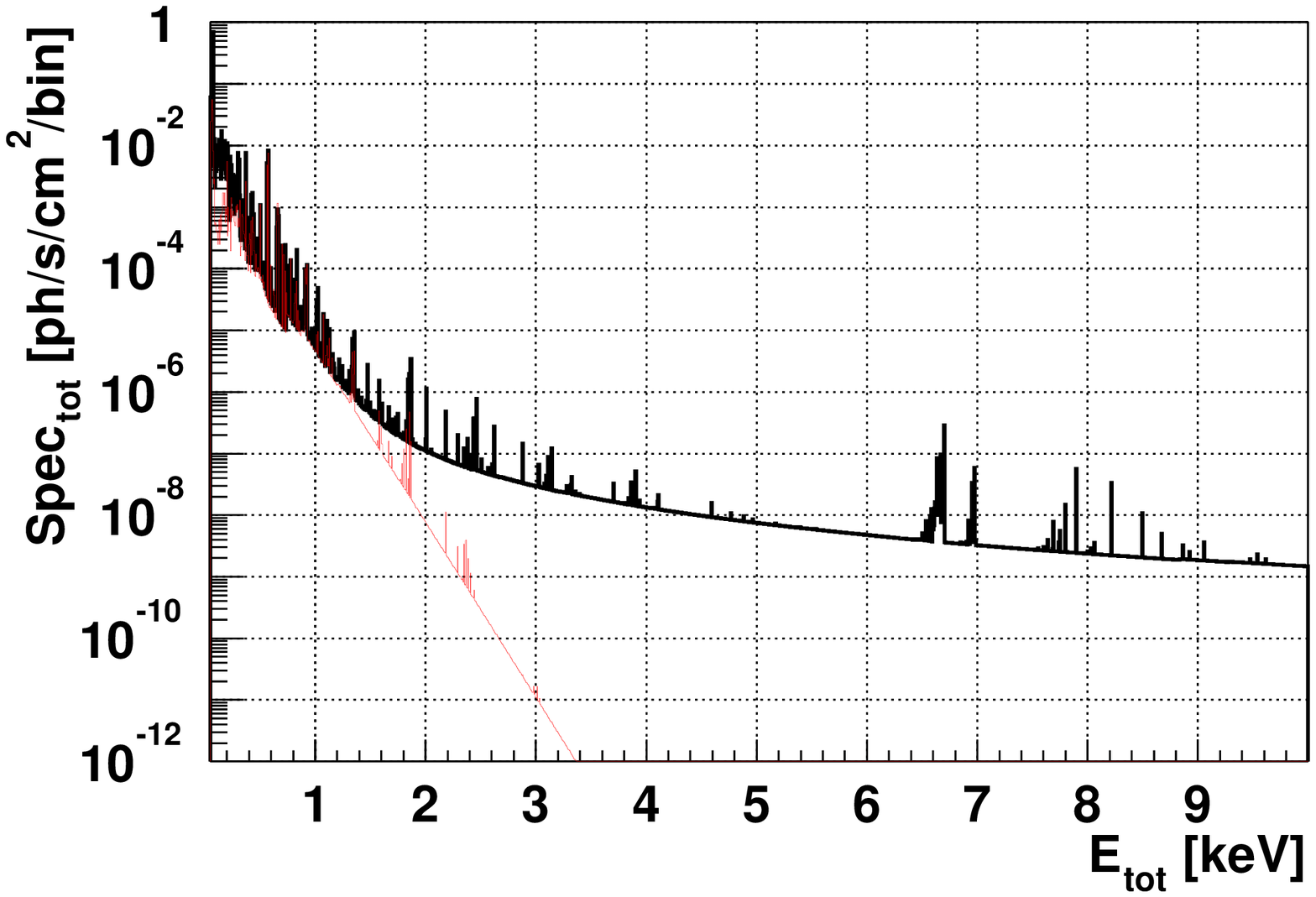}}
\caption[]{The solar X-ray spectrum reconstructed from the {\it emission measure}
distribution (EM(T)) for the non-flaring Sun at the solar minimum
\cite{orlando}. A thermal component of $\sim 1.8$ MK is also shown (thin line in
red). Bin size = 6.1 eV.

(EM(T) is approximately the product of the
square of the electron density with the emitting volume as a function of
temperature, i.e.  EM(T) = $\rho_e^2(T)\cdot V(T)$
\cite{drake4321}).}
\end{figure}


\begin{thebibliography}{99}
\bibitem{morphology} The morphology of the transition region remains
 unsolved \cite{warren9zzz,wikstol9zzz}. The question is not whether such a thin
 interface exist, but whether it contributes substatially to the bright emission
 seen in transition region lines on the solar disk (see also ref.\cite{doschek9zzz}).
\bibitem{giomatarisx}  Measurement performed in Saclay in collaboration 
 with the group of Y.~Giomataris, using a Micromegas chamber 
 \cite{giomataris} 
 previously exposed to a beam. Charged particles were rejected.
\end{thebibliography}
\end{document}